\newif\ifarxiv
\arxivtrue 

\documentclass[runningheads]{llncs}

\usepackage[T1]{fontenc}

\usepackage{amsmath}
\usepackage{amssymb}
\usepackage{enumitem}
\usepackage{mathtools}
\usepackage{xcolor,colortbl}
\usepackage{xfrac}
\usepackage{tabularx,tabu}
\usepackage{enumerate}
\usepackage{stmaryrd}
\usepackage{algorithm,algorithmicx}
\usepackage{xspace}
\usepackage{xfrac}

\usepackage{caption}
\usepackage{subcaption}

\usepackage[noend]{algpseudocode}
\usepackage{bm}
\usepackage[b]{esvect}
 \usepackage{tikz,pgf}
 \usetikzlibrary{positioning,calc,backgrounds,automata}
 \usepackage[bb=boondox]{mathalfa}
 \usepackage{wrapfig}
 \usepackage{siunitx}
 \usepackage{array}
 \usepackage{multirow}

 \newtheorem{assumption}{Assumption}
 \newtheorem{prob}{Problem}

\newcommand{\tup}[1]{\left(#1 \right)}
\newcommand{\set}[1]{\lbrace #1\rbrace}
\newcommand{\pr}{\mathbb{P}}

\newcommand{\expe}{\mathbb{E}}
\newcommand{\sem}[1]{\llbracket #1\rrbracket}
\newcommand{\distr}{\mathcal{D}}

\newcommand{\interval}[2]{[#1 \pm #2]}

\newcommand{\bernoulli}[1]{\mathit{Bernoulli}(#1)}

\newcommand{\simplex}{\Delta}
 \newcommand{\twodots}{\mathinner {\ldotp \ldotp}}
 \newcommand{\range}[1]{[1\twodots #1]}
 
 \newcommand\Myperm[2][^n]{\prescript{#1\mkern-2.5mu}{}P_{#2}}

\newcommand{\seq}[1]{\vv{#1}}

\newcommand{\algfreqdivfree}{\texttt{FreqMonitorDivFree}\xspace}
\newcommand{\algfreq}{\texttt{FreqMonitor}\xspace}

\newcommand{\mc}{\mathcal{M}}
\newcommand{\Q}{Q}

\newcommand{\trm}{M}
\newcommand{\paths}[1]{\mathit{Paths}(#1)}
\newcommand{\initd}{\pi}

\newcommand{\prd}{\ensuremath{p_{\theta}}}
\newcommand{\mcset}{\ensuremath{\mathbb{M}}}

\newcommand{\monitor}{\mathcal{A}}
\newcommand{\verdict}{\Lambda}
\newcommand{\dom}{\mathit{Dom}}

\newcommand{\depends}{\ensuremath{\mathit{Dep}}}
\newcommand{\transrel}{T}

\algnewcommand{\IfThenElse}[3]{
  \State \algorithmicif\ #1\ \algorithmicthen\ #2\ \algorithmicelse\ #3}
  
\newcommand{\bpr}[1]{\mathbb{P}_{#1}}

\newcommand{\lp}[2]{v_{#1#2}}
\newcommand{\mcm}{\matr{M}}
\newcommand{\mcmr}{M}

\newcommand{\para}{\theta}

\newcommand{\nconst}{\mathcal{N}}
\newcommand{\likely}{\mathcal{L}}

\newcommand{\indi}{I}
\newcommand{\indj}{J}

\newcommand{\mlp}{D}

\newcommand{\isum}[1]{\left\langle #1\right\rangle}

\newcommand{\matr}[1]{#1}

\newcommand{\iinter}[2]{[#1..#2]}

\newcommand{\mfact}[3]{#3!_{#2}^{#1}}
\newcommand{\ufact}[2]{\mathcal{H}_{#1}\left(#2\right)}

\newcommand{\uifact}[4]{\mathcal{H}^{#1#2}_{#3}\left(#4\right)}

\newcommand{\matru}[2]{\matr{\delta^{#1#2}}}

\newcommand{\beval}[2]{#1(#2)}

\newcommand{\uf}{\mathcal{H}}

\newcommand{\activ}{\ensuremath{\mathsf{active}}}

\begin{document}
\title{Monitoring Algorithmic Fairness\thanks{This work is supported
by the European Research Council under Grant No.: \mbox{ERC-2020-AdG
101020093.}}}

\author{Thomas A.\ Henzinger \orcidID{0000-0002-2985-7724} \and\\
		Mahyar Karimi \orcidID{0009-0005-0820-1696} \and\\
		Konstantin Kueffner \orcidID{0000-0001-8974-2542} \and\\
		Kaushik Mallik \orcidID{0000-0001-9864-7475}}
		
\authorrunning{Henzinger et al.}
		
\institute{Institute of Science and Technology Austria (ISTA), Klosterneuburg, Austria.\\
	\email{\{tah, mahyar.karimi, konstantin.kueffner, kaushik.mallik\}@ist.ac.at}}

\setlength{\textfloatsep}{12pt plus 3.0pt minus 3.0pt}

	\maketitle
	
	\begin{abstract}
		Machine-learned systems are in widespread use for making decisions about humans, and it is important that they are \emph{fair}, i.e., not biased against individuals based on sensitive attributes.
		We present runtime verification of algorithmic fairness for systems whose models are unknown, but are assumed to have a Markov chain structure.
		We introduce a specification language that can model many common algorithmic fairness properties, such as demographic parity, equal opportunity, and social burden.
		We build monitors that observe a long sequence of events as generated by a given system, and output, after each observation, a quantitative estimate of how fair or biased the system was on that run until that point in time.
		The estimate is proven to be correct modulo a variable error bound and a given confidence level, where the error bound gets tighter as the observed sequence gets longer.
		Our monitors are of two types, and use, respectively, frequentist and Bayesian statistical inference techniques. 
		While the frequentist monitors compute estimates that are objectively correct with respect to the ground truth, the Bayesian monitors compute estimates that are correct subject to a given prior belief about the system's model. 
		Using a prototype implementation, we show how we can monitor if a bank is fair in giving loans to applicants from different social backgrounds, and if a college is fair in admitting students while maintaining a reasonable financial burden on the society.
		Although they exhibit different theoretical complexities in certain cases, in our experiments, both frequentist and Bayesian monitors took less than a millisecond to update their verdicts after each observation.
	\end{abstract}


\section{Introduction}
\label{sec:intro}

Runtime verification complements traditional static verification techniques, by offering lightweight solutions for checking properties based on a single, possibly long execution trace of a given system \cite{bartocci2018lectures}.
We present new runtime verification techniques for the problem of bias detection in decision-making software.
The use of software for making critical decisions about humans is a growing trend; example areas include judiciary \cite{chouldechova2017fair,dressel2018accuracy}, policing \cite{ensign2018runaway,lum2016predict}, banking \cite{liu2018delayed}, etc.
It is  important that these software systems are unbiased towards the protected attributes of humans, like gender, ethnicity, etc.
However, they have often shown biases in their decisions in the past \cite{dressel2018accuracy,lahoti2019ifair,obermeyer2019dissecting,scheuerman2019computers,seyyed2020chexclusion}.
While there are many approaches for mitigating biases before deployment \cite{dressel2018accuracy,lahoti2019ifair,obermeyer2019dissecting,scheuerman2019computers,seyyed2020chexclusion}, recent runtime verification approaches \cite{albarghouthi2019fairness,henzinger2023runtime} offer a new complementary tool to oversee \emph{algorithmic fairness} in AI and machine-learned decision makers during deployment.

To verify algorithmic fairness at runtime, the given decision-maker is treated as a \emph{generator} of events with an unknown model.
The goal is to algorithmically design lightweight but rigorous \emph{runtime monitors} against quantitative formal specifications. 
The monitors observe  a long stream of events and, after each observation, output a quantitative, statistically sound estimate of how fair or biased the generator was until that point in time.
While the existing approaches \cite{albarghouthi2019fairness,henzinger2023runtime} considered only sequential decision making models and built monitors from the frequentist viewpoint in statistics, we allow the richer class of Markov chain models and present monitors from both the frequentist and the Bayesian statistical viewpoints.

Monitoring algorithmic fairness involves on-the-fly statistical estimations, a feature that has not been well-explored in the traditional runtime verification literature.
As far as the algorithmic fairness literature is concerned, the existing works are mostly \emph{model-based}, and either minimize decision biases of machine-learned systems at \emph{design-time} (i.e., pre-processing) \cite{kamiran2012data,zemel2013learning,berk2017convex,zafar2019fairness}, or verify their absence at \emph{inspection-time} (i.e., post-processing) \cite{hardt2016equality}.
In contrast, we verify algorithmic fairness at \emph{runtime}, and do not require an explicit model of the generator.
On one hand, the model-independence makes the monitors trustworthy, and on the other hand, it complements the existing model-based static analyses and design techniques, which are often insufficient due to partially unknown or imprecise models of systems in real-world environments.

We assume that the sequences of events generated by the generator can be modeled as sequences of states visited by a finite unknown Markov chain. 
This implies that the generator is well-behaved and the events follow each other according to some fixed probability distributions.
Not only is this assumption satisfied by many machine-learned systems (see Sec.~\ref{sec:motivating examples} for examples), it also provides just enough structure to lay the bare-bones foundations for runtime verification of algorithmic fairness properties. 
We emphasize that we do not require knowledge of the transition probabilities of the underlying Markov chain.

We propose a new specification language, called the Probabilistic Specification Expressions (PSEs), which can formalize a majority of the existing algorithmic fairness properties in the literature, including demographic parity \cite{dwork2012fairness}, equal opportunity \cite{hardt2016equality}, disparate impact \cite{feldman2015certifying}, etc. 
Let $Q$ be the set of events.
Syntactically, a PSE is a restricted arithmetic expression over the (unknown) transition probabilities of a Markov chain with the state space $Q$.
Semantically, a PSE $\varphi$ over $Q$ is a function that maps every Markov chain $M$ with the state space $Q$ to a real number, and the value $\varphi(M)$ represents the degree of fairness or bias (with respect to $\varphi$) in the generator $M$. 
Our monitors observe a long sequence of events from $Q$, and after each observation, compute a statistically rigorous estimate of $\varphi(M)$ with a PAC-style error bound for a given confidence level.
As the observed sequence gets longer, the error bound gets tighter.

Algorithmic fairness properties that are expressible using PSEs are quantitative refinements of the traditional qualitative fairness properties studied in formal methods.
For example, a qualitative fairness property may require that if a certain event $A$ occurs infinitely often, then another event $B$ should follow infinitely often.
In particular, a coin is qualitatively fair if infinitely many coin tosses contain both infinitely many heads and infinitely many tails.
In contrast, the coin will be algorithmically fair (i.e., unbiased) if approximately half of the tosses come up heads.
Technically, while qualitative weak and strong fairness properties are $\omega$-regular, the algorithmic fairness properties are statistical and require counting.
Moreover, for a qualitative fairness property, the satisfaction or violation cannot be established based on a finite prefix of the observed sequence.
In contrast, for any given finite prefix of observations, the value of an algorithmic fairness property can be estimated using statistical techniques, assuming the future behaves statistically like the past (the Markov assumption).

As our main contribution, we present two different monitoring algorithms, using tools from frequentist and Bayesian statistics, respectively.
The central idea of the \emph{frequentist monitor} is that the probability of every transition of the monitored Markov chain $M$ can be estimated using the fraction of times the transition is taken per visit to its source vertex.
Building on this, we present a practical implementation of the frequentist monitor that can estimate the value of a given PSE from an observed finite sequence of states. 
For the coin example, after every new toss, the frequentist monitor will update its estimate of probability of seeing heads by computing the fraction of times the coin came up heads so far, and then by using concentration bounds to find a tight error bound for a given confidence level.
On the other hand, the central idea of the \emph{Bayesian monitor} is that we begin with a prior belief about the transition probabilities of $M$, and having seen a finite sequence of observations, we can obtain an updated posterior belief about $M$.
For a given confidence level, the output of the monitor is computed by applying concentration inequalities to find a tight error bound around the mean of the posterior belief.
For the coin example, the Bayesian monitor will begin with a prior belief about the degree of fairness, and, after observing the outcome of each new toss, will compute a new posterior belief.
If the prior belief agrees with the true model with a high probability, then the Bayesian monitor's output converges to the true value of the PSE more quickly than the frequentist monitor.
In general, both monitors can efficiently estimate more complicated PSEs, such as the ratio and the squared difference of the probabilities of heads of two different coins.
The choice of the monitor for a particular application depends on whether an objective or a subjective evaluation, with respect to a given prior, is desired.

Both frequentist and Bayesian monitors use registers (and counters as a restricted class of registers) to keep counts of the relevant events and store the intermediate results.
If the size of the given PSE is $n$, then, in theory, the frequentist monitor uses $\mathcal{O}(n^42^n)$ registers and computes its output in $\mathcal{O}(n^42^n)$ time after each new observation, whereas the Bayesian monitor uses $\mathcal{O}(n^22^n)$ registers and computes its output in $\mathcal{O}(n^22^n)$ time after each new observation.
The computation time and the required number of registers get drastically reduced to $\mathit{O}(n^2)$ for the frequentist monitor with PSEs that contain up to one division operator, and for the Bayesian monitor with polynomial PSEs (possibly having negative exponents in the monomials). 
This shows that under given circumstances, one or the other type of the monitor can be favorable computation-wise.
These special, efficient cases cover many algorithmic fairness properties of interest, such as demographic parity and equal opportunity.

Our experiments confirm that our monitors are fast in practice.
Using a prototype implementation in Rust, we monitored a couple of decision-making systems adapted from the literature.
In particular, we monitor if a bank is fair in lending money to applicants from different demographic groups \cite{liu2018delayed}, and if a college is fair in admitting students without creating an unreasonable financial burden on the society \cite{milli2019social}.
In our experiments, both monitors took, on an average, less than a millisecond to update their verdicts after each observation, and only used tens of internal registers to operate, thereby demonstrating their practical usability at runtime.

In short, we advocate that runtime verification introduces a new set of tools in the area of algorithmic fairness, using which we can monitor biases of deployed AI and machine-learned systems in real-time.
While existing monitoring approaches only support sequential decision making problems and use only the frequentist statistical viewpoint, we present monitors for the more general class of Markov chain system models using both frequentist and Bayesian statistical viewpoints.

\ifarxiv
	All proofs can be found in the appendix.
\else
	All proofs can be found in the longer version of the paper \cite{??}.
\fi

\subsection{Motivating Examples}
\label{sec:motivating examples}

We first present two real-world examples from the algorithmic fairness literature to motivate the problem; these examples will later be used to illustrate the technical developments.

\smallskip
\noindent\textbf{The lending problem \cite{liu2018delayed}:}
Suppose a bank lends money to individuals based on certain attributes, like credit score, age group, etc.
The bank wants to maximize profit by lending money to only those who will repay the loan in time---called the ``true individuals.''
There is a sensitive attribute (e.g., ethnicity) classifying the population into two groups $g$ and $\overline{g}$.
The bank will be considered fair (in lending money) if its lending policy is independent of an individual's membership in $g$ or $\overline{g}$.
Several \emph{group fairness} metrics from the literature are relevant in this context.
\emph{Disparate impact} \cite{feldman2015certifying} quantifies the \emph{ratio} of the probability of an individual from $g$ getting the loan to the probability of an individual from $\overline{g}$ getting the loan, which should be close to $1$ for the bank to be considered fair.
\emph{Demographic parity} \cite{dwork2012fairness} quantifies the \emph{difference} between the probability of an individual from $g$ getting the loan and the probability of an individual from $\overline{g}$ getting the loan, which should be close to $0$ for the bank to be considered fair.
\emph{Equal opportunity} \cite{hardt2016equality} quantifies the \emph{difference} between the probability of a \emph{true} individual from $g$ getting the loan and the probability of a \emph{true} individual from $\overline{g}$ getting the loan, which should be close to $0$ for the bank to be considered fair.
A discussion on the relative merit of various different algorithmic fairness notions is out of scope of this paper, but can be found in the literature \cite{wachter2020bias,kleinberg2016inherent,corbett2017algorithmic,dwork2018individual}.
We show how we can monitor whether a given group fairness criteria is fulfilled by the bank, by observing a sequence of lending decisions.

\smallskip
\noindent\textbf{The college admission problem \cite{milli2019social}:}
Consider a college that announces a cutoff of grades for admitting students through an entrance examination.
Based on the merit, every truly qualified student belongs to group $g$, and the rest to group $\overline{g}$.
Knowing the cutoff, every student can choose to invest a sum of money---proportional to the gap between the cutoff and their true merit---to be able to reach the cutoff, e.g., by taking private tuition classes.
On the other hand, the college's utility is in minimizing admission of students from $\overline{g}$, which can be accomplished by raising the cutoff to a level that is too expensive to be achieved by the students from $\overline{g}$ and yet easy to be achieved by the students from $g$.
The \emph{social burden} associated to the college's cutoff choice is the expected expense of every student from $g$, which should be close to $0$ for the college to be considered fair (towards the society).
We show how we can monitor the social burden, by observing a sequence of investment decisions made by the students from $g$.

\subsection{Related Work}
There has been a plethora of work on algorithmic fairness from the machine learning standpoint  \cite{mehrabi2021survey,dwork2012fairness,hardt2016equality,kusner2017counterfactual,kearns2018preventing,sharifi2019average,bellamy2019ai,wexler2019if,bird2020fairlearn,zemel2013learning,jagielski2019differentially,konstantinov2022fairness}.
In general, these works improve algorithmic fairness through de-biasing the training dataset (pre-processing), or through incentivizing the learning algorithm to make fair decisions (in-processing), or through eliminating biases from the output of the machine-learned model (post-processing).
All of these are interventions in the design of the system, whereas our monitors treat the system as already deployed.

Recently, formal methods-inspired techniques have been used to guarantee algorithmic fairness through the verification of a learned model  \cite{albarghouthi2017fairsquare,bastani2019probabilistic,sun2021probabilistic,ghosh2020justicia,meyer2021certifying}, and enforcement of robustness \cite{john2020verifying,balunovic2021fair,ghosh2021algorithmic}.
All of these works verify or enforce algorithmic fairness \emph{statically} on all runs of the system with high probability.
This requires certain knowledge about the system model, which may not be always available. 
Our runtime monitor dynamically verifies whether the current run of an opaque system is fair.

Our frequentist monitor is closely related to the novel work of Albarghouthi et al.~\cite{albarghouthi2019fairness}, where the authors build a programming framework that allows runtime monitoring of algorithmic fairness properties on programs.
Their monitor evaluates the algorithmic fairness of repeated ``single-shot'' decisions made by machine-learned functions on a sequence of samples drawn from an underlying unknown but fixed distribution, which is a special case of our more general Markov chain  model of the generator.
They do not consider the Bayesian point of view.
Moreover, we argue and empirically show in Sec.~\ref{sec:frequentist:background} that our frequentist approach produces significantly tighter statistical estimates than their approach on most PSEs.
On the flip side, their specification language is more expressive, in that they allow atomic variables for expected values of events, which is useful for specifying individual fairness criteria \cite{dwork2012fairness}.
We only consider group fairness, and leave individual fairness as part of future research. 
Also, they allow logical operators (like boolean connectives) in their specification language.
However, we obtain tighter statistical estimates for the core arithmetic part of algorithmic fairness properties (through PSEs), and point out that we can deal with logical operators just like they do in a straightforward manner.

Shortly after the first manuscript of this paper was written, we published a separate work for monitoring long-run fairness in sequential decision making problems, where the feature distribution of the population may dynamically change due to the actions of the individuals \cite{henzinger2023runtime}.
Although this other work generalizes our current paper in some aspects (support for dynamic changes in the model), it only allows sequential decision making models (instead of Markov chains) and does not consider the Bayesian monitoring perspective.

There is a large body of research on monitoring, though the considered properties are mainly temporal \cite{stoller2011runtime,junges2021runtime,faymonville2017real,maler2004monitoring,donze2010robust,bartocci2018specification,baier2003ctmc}.
Unfortunately, these techniques do not directly extend to monitoring algorithmic fairness, since checking algorithmic fairness requires statistical methods, which is beyond the limit of finite automata-based monitors used by the classical techniques.
Although there are works on quantitative monitoring that use richer types of monitors (with counters/registers like us) \cite{finkbeiner2002collecting,henzinger2020monitorability,otop2019quantitative,henzinger2021quantitative}, the considered specifications do not easily extend to statistical properties like algorithmic fairness.
One exception is the work by Ferr{\`e}re et al.~\cite{ferrere2019monitoring}, which monitors certain statistical properties, like mode and median of a given sequence of events. 
Firstly, they do not consider algorithmic fairness properties.
Secondly, their monitors' outputs are correct only as the length of the observed sequence approaches infinity (asymptotic guarantee), whereas our monitors' outputs are \emph{always} correct with high confidence (finite-sample guarantee), and the precision gets better for longer sequences.

Although our work uses similar tools as used in statistical verification \cite{ashok2019pac,younes2002probabilistic,clarke2011statistical,david2013optimizing,agha2018survey}, the goals are different. 
In traditional statistical verification, the system's runs are chosen probabilistically, and it is verified if any run of the system satisfies a boolean property with a  certain probability.
For us, the run is given as input to the monitor, and it is this run that is verified against a quantitative algorithmic fairness property with statistical error bounds.
To the best of our knowledge, existing works on statistical verification do not consider algorithmic fairness properties.


\section{Preliminaries}
For any alphabet $\Sigma$, the notation $\Sigma^*$ represents the set of all finite words over $\Sigma$.
We write $\mathbb{R}$, $\mathbb{N}$, and $\mathbb{N}^+$ to denote the sets of real numbers, natural numbers (including zero), and positive integers, respectively.
For a pair of real (natural) numbers $a,b$ with $a < b$, we write $[a,b]$ ($[a\twodots b]$) to denote the set of all real (natural) numbers between and including $a$ and $b$. 
For a given $c,r\in \mathbb{R}$, we write $\interval{c}{r}$ to denote the set $[c-r,c+r]$.
For simpler notation, we will use $|\cdot|$ to denote both the cardinality of a set and the absolute value of a real number, whenever the intended use is clear.

For a given vector $v\in \mathbb{R}^n$ and a given $m\times n$ real matrix $M$, for some $m,n$, we write $v_i$ to denote the $i$-th element of $v$ and write $M_{ij}$ to denote the element at the $i$-th row and the $j$-th column of $M$.
For a given $n\in \mathbb{N}^+$, a \emph{simplex} is the set of vectors $\simplex(n)\coloneqq \set{x\in  [0,1]^{n+1} \mid \sum_{i=1}^{n+1} x_i = 1}$.
Notice that the dimension of $\simplex(n)$ is $n+1$ (and not $n$), a convention that is standard due to the interpretation of $\simplex(n)$ as the $n+1$ vertices of an $n$-dimensional polytope.
A \emph{stochastic matrix} of dimension $m\times m$ is a matrix whose every row is in $\simplex(m-1)$, i.e.\ $M\in \simplex(m-1)^m$.
Random variables 
\ifarxiv
	(see App.~\ref{app:random variables})
\else
\fi 
will be denoted using uppercase symbols from the Latin alphabet (e.g.\ $X$), while the associated outcomes will be denoted using lowercase font of the same symbol ($x$ is an outcome of $X$).
We will interchangeably use the expected value $\expe(X)$ and the mean $\mu_X$ of $X$.
For a given set $S$, define $\distr(S)$ as the set of every random variable---called a \emph{probability distribution}\footnote{An alternate commonly used definition of probability distribution is directly in terms of the probability measure induced over $S$, instead of through the random variable.}%
---with set of outcomes being $2^S$.
A Bernoulli random variable that produces ``$1$'' (the alternative is ``$0$'') with probability $p$ is written as $\bernoulli{p}$. 

\subsection{Markov chains as randomized generators of events}
We use finite Markov chains as sequential randomized generators of events.
A (finite) Markov chain $\mc$ is a triple $ (\Q,\trm,\initd)$, where $\Q=\range{N}$ is a set of states for a finite $N$, $\trm\in \simplex(N-1)^N$ is a stochastic matrix called the transition probability matrix, and $\initd\in \distr(\Q)$ is the distribution over initial states.
We often refer to a pair of states $(i,j)\in \Q\times\Q$ as an \emph{edge}.
The Markov chain $\mc$ generates an infinite sequence of random variables $X_0=\initd,X_1,\ldots$, with $X_i\in \distr(\Q)$ for every $i$, such that the Markov property is satisfied:
	$\pr(X_{n+1} = i_{n+1} \mid X_0=i_0,\ldots,X_n=i_n) = \pr(X_{n+1}=i_{n+1} \mid X_n=i_n)$,
which is $\trm_{i_ni_{n+1}}$ in our case.
A finite \emph{path} $\seq{x}=x_0,\ldots,x_n$ of $\mc$ is a finite word over $\Q$ such that for every $t\in [0;n]$, $\pr(X_t=x_t)>0$.
Let $\paths{\mc}$ be the set of every finite path of $\mc$.

We use Markov chains to model the probabilistic interaction between a machine-learned decision maker with its environment.
Intuitively, the Markov assumption on the model puts the restriction that the decision maker does not change over time, e.g., due to retraining.

\begin{figure}
		\centering
		\vspace*{-0.5cm}
		\begin{subfigure}{0.50\textwidth}
			\centering
			\def\H{1}
			\def\W{1.7}
			\begin{tikzpicture}[every node/.style={scale=0.75},scale=0.8,baseline=(init.base)]
				\node[state] (init)    at (0,0)        {\textit{init}};
				\node[state] (g-c)     at (\W^1.5, 0)  {$  \overline{y}$};
				\node[state] (g-c-A)   at (\W^2, \H)   {$ gy$};
				\node[state] (g-c-R)   at (\W^2,-\H)   {$\overline{g}y$};
				\node[state] (g-c-A-S) at (\W^2.9, \H) {$ z$};
				\node[state] (g-c-A-F) at (\W^2.9,-\H) {$\overline{z}$};

				\path[->] 
				  (g-c) edge (init)
					(g-c-A)   edge (g-c-A-S) edge (g-c-A-F)
					(g-c-A-S) edge[out=135,in=60] (init)
					(g-c-A-F) edge[out=-135,in=-60] (init)
					(g-c-R)   edge (g-c-A-S) edge	(g-c-A-F);
					
				\node[state]	(env-ch-1)	at	(\W^0.7,\H)		{$g$};
				\node[state]	(env-ch-2)	at	(\W^0.7,-\H)	{$\overline{g}$};
				\path[->]
					(env-ch-1) edge (g-c-A)	edge (g-c)
					(env-ch-2) edge (g-c-R)	edge (g-c)
					(init) edge (env-ch-1) edge	(env-ch-2);
			\end{tikzpicture}
		\end{subfigure}
		\quad
		\begin{subfigure}{0.45\textwidth}
			\centering
			\def\H{1.3}
			\def\W{1.6}
			\begin{tikzpicture}[every node/.style={scale=0.75},scale=0.8,baseline=(init.base)]
				\node[state] (init) at (0,0) {\textit{init}};
					
				\node[state]	(env-ch-1)	at (\W^0.7,\H)	 {$g$};
				\node[state]	(env-ch-2)	at (\W^0.7,-\H)	 {$\overline{g}$};
				\node[state]  (merit-0)   at (\W^2.2, \H)  {$0$};
				\node[state]	(merit-1)	  at (\W^2.2,0.33) {$1$};
				\node       	(merit-2)	  at (\W^2.2,-0.5) {$\vdots$};
				\node[state]	(merit-3)	  at (\W^2.2,-\H)	 {$N$};
				\path[->]
					(env-ch-1) edge (merit-0)
					(init) edge	(env-ch-1) edge (env-ch-2)
					(merit-1)	edge (init)
					(merit-0)	edge (init)
					(merit-3)	edge (init)
					(env-ch-1) edge	(merit-1)	edge (merit-3)
					(env-ch-2) edge (merit-0)
					(env-ch-2) edge (merit-1)
					(env-ch-2) edge (merit-3);
 			\end{tikzpicture}
		\end{subfigure}
		\vspace*{-0.25cm}
		\caption{Markov chains for the lending and the college-admission examples.
		(\textbf{left})~The lending example: The state $\textit{init}$ denotes the initiation of the sampling, and the rest represent the selected individual, namely, $g$ and $\overline{g}$ denote the two groups, $(gy)$ and $(\overline{g}y)$ denote that the individual is respectively from group $g$ and group $\overline{g}$ and the loan was granted, $\overline{y}$ denotes that the loan was refused,	and $z$ and $\overline{z}$ denote whether the loan was repaid or not.
		(\textbf{right})~The college admission example: The state \textit{init} denotes the initiation of the sampling, the states $g,\overline{g}$ represent the group identity of the selected candidate, and the states $\set{0,\ldots,N}$ represent the amount of money invested by a truly eligible candidate.}
		\label{fig:markov chain examples}
\end{figure}
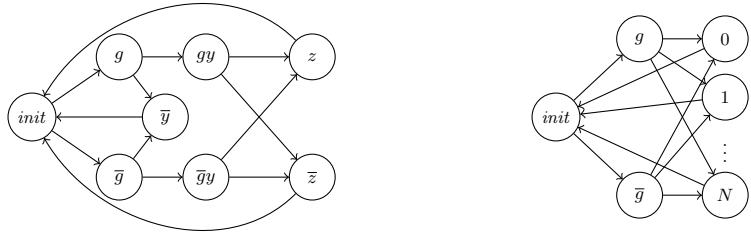

In Fig.~\ref{fig:markov chain examples} we show the Markov chains for the lending and the college admission examples from Sec.~\ref{sec:motivating examples}.
The Markov chain for the lending example captures the sequence of loan-related probabilistic events, namely, that a loan applicant is randomly sampled and the group information ($g$ or $\overline{g}$) is revealed, a probabilistic decision is made by the decision-maker and either the loan was granted ($gy$ or $\overline{g}y$, depending on the group) or refused ($\overline{y}$), and if the loan is granted then with some probabilities it either gets repaid ($z$) or defaulted ($\overline{z}$).
The Markov chain for the college admission example captures the sequence of admission events, namely, that a candidate is randomly sampled and the group is revealed ($g,\overline{g}$), and when the candidate is from group $g$ (truly qualified) then the amount of money invested for admission is also revealed.

\subsection{Randomized register monitors}
Randomized register monitors, or simply monitors, are adapted from the (deterministic) polynomial monitors of Ferr\`ere et al.\ \cite{ferrere2018theory}.
Let $R$ be a finite set of integer variables called registers.
A function $v\colon R\to \mathbb{N}$ assigning concrete value to every register in $R$ is called a valuation of $R$.
Let $\mathbb{N}^R$ denote the set of all valuations of $R$.
Registers can be read and written according to relations in the signature $S=\langle 0,1,+,-,\times,\div,\leq \rangle$.
We consider two basic operations on registers:
\begin{itemize}[noitemsep,topsep=0pt]
	\item A \emph{test} is a conjunction of atomic formulas over $S$ and their negation;
	\item An \emph{update} is a mapping from variables to terms over $S$.
\end{itemize}
We use $\Phi(R)$ and $\Gamma(R)$ to respectively denote the set of tests and updates over $R$.
\emph{Counters} are special registers with a restricted signature $S=\langle 0,1,+,-,\leq \rangle$.

\begin{definition}[Randomized register monitor]
	A randomized register monitor is a tuple $\tup{\Sigma,\Lambda,R,\lambda,\transrel}$ where 
	$\Sigma$ is a finite input alphabet, 
	$\verdict$ is an output alphabet, 
	$R$ is a finite set of registers,
	$\lambda\colon \mathbb{N}^R\to \verdict$ is an output function, and
	$\transrel\colon \Sigma\times\Phi(R)\to \distr(\Gamma(R))$ is the randomized transition function such that
	for every $\sigma\in \Sigma$ and for every valuation $v\in \mathbb{N}^R$, there exists a unique $\phi\in \Phi(R)$ with $v\models\phi$ and $\transrel(\sigma,\phi)\in\distr(\Gamma(R))$.
	A deterministic register monitor is a randomized register monitor for which $\transrel(\sigma,\phi)$ is a Dirac delta distribution, if it is defined.
\end{definition}

A \emph{state} of a monitor $\monitor$ is a valuation of its registers $v\in \mathbb{N}^R$.
The monitor $\monitor$ \emph{transitions} from state $v$ to a \emph{distribution} over states given by the random variable $Y=\transrel(\sigma,\phi)$ on input $\sigma\in \Sigma$ if there exists $\phi$ such that $v\models \phi$.
Let $\gamma$ be an outcome of $Y$ with $\pr(Y=\gamma)>0$, in which case the registers are updated as $v'(x)=v(\gamma(x))$ for every $x\in R$, and the respective concrete transition is written as $v\xrightarrow{\sigma} v'$.
A \emph{run} of $\monitor$ on a word $w_0\ldots w_n\in \Sigma^*$ is a sequence of concrete transitions $v_0\xrightarrow{w_0} v_1\xrightarrow{w_1}\ldots\xrightarrow{w_n} v_{n+1}$. 
The probabilistic transitions of $\monitor$ induce a probability distribution over the sample space of finite runs of the monitor, denoted $\widehat{\pr}(\cdot)$.
For a given finite word $w\in \Sigma^*$, the \emph{semantics} of the monitor $\monitor$ is given by a random variable $\sem{\monitor}(w)\coloneqq \lambda(Y)$ inducing the probability measure $\pr_\monitor$, where $Y$ is the random variable representing the distribution over the final state in a run of $\monitor$ on the word $w$, i.e., $\pr_\monitor(Y=v) \coloneqq \widehat{\pr}(\set{r=r_0\ldots r_m\in \Sigma^*\mid r \text{ is a run of } \monitor \text{ on } w \text{ and } r_m=v})$.

\medskip
\noindent\textbf{Example: A monitor for detecting the (unknown) bias of a coin.}
We present a simple deterministic monitor that computes a PAC estimate of the bias of an unknown coin from a sequence of toss outcomes, where the outcomes are denoted as ``$h$'' for heads and ``$t$'' for tails.
The input alphabet is the set of toss outcomes, i.e., $\Sigma=\set{h,t}$, 
the output alphabet is the set of every bias intervals, i.e., $\Gamma = \set{[a,b]\mid 0\leq a<b\leq 1}$, 
the set of registers is $R=\set{r_n,r_h}$, where $r_n$ and $r_h$ are counters counting the total number of tosses and the number of heads, respectively, and
the output function $\lambda$ maps every valuation of $r_n,r_h$ to an interval estimate of the bias that has the form $\lambda \equiv \sfrac{v(r_h)}{v(r_n)} \pm \varepsilon(r_n,\delta)$, where $\delta\in [0,1]$ is a given upper bound on the probability of an incorrect estimate and $\varepsilon(r_n,\delta)$ is the estimation error computed using PAC analysis.
For instance, after observing a sequence of $67$ tosses with $36$ heads, the values of the registers will be $v(r_n)=67$ and $v(r_h)=36$, and the output of the monitor will be $\lambda(67,36) = \sfrac{36}{67} \pm \varepsilon(n,\delta)$ for some appropriate $\varepsilon(\cdot)$.
Now, suppose the next input to the monitor is $h$, in which case the monitor's transition is given as $T(h,\cdot) = (r_n+1,r_h+1)$, which updates the registers to the new values $v'(r_n)=67+1=68$ and $v'(r_h)=36+1=37$.
For this example, the tests $\Phi(R)$ over the registers are redundant, but they can be used to construct monitors for more complex properties.

\section{Algorithmic Fairness Specifications and\\ Problem Formulation}

\subsection{Probabilistic Specification Expressions}
\label{sec:problem:probability properties}

To formalize algorithmic fairness properties, like the ones in Sec.~\ref{sec:motivating examples}, we introduce \emph{probabilistic specification expressions} (PSE). 
A PSE $\varphi$ over a given finite set $\Q$ is an algebraic expression with some restricted set of operations that uses variables labeled $v_{ij}$ with $i,j\in\Q$ and whose domains are the real interval $[0,1]$.
The syntax of $\varphi$ is:
\begin{subequations}\label{equ:syntax property}
\begin{alignat}{2}
	\xi &\Coloneqq  v\in \set{v_{ij}}_{i,j\in\Q} \ | \ \xi\cdot\xi \ | \ 1\div\xi, \label{equ:syntax division and subtraction-free probability prop.}\\
	\varphi &\Coloneqq \kappa \in \mathbb{R} \ | \ \xi \ | \  \varphi + \varphi \ | \ \varphi - \varphi \ | \ \varphi\cdot\varphi \ | \ (\varphi), \label{equ:syntax probability prop.}
\end{alignat}
\end{subequations}
where $\set{v_{ij}}_{i,j\in\Q}$ are the variables with domain $[0,1]$ and $\kappa$ is a constant.
The expression $\xi$ in \eqref{equ:syntax division and subtraction-free probability prop.} is called a \emph{monomial} and is simply a product of powers of variables with integer exponents.
A \emph{polynomial} is a weighted sum of monomials with constant weights.\footnote{Although monomials and polynomials usually only have positive exponents, we take the liberty to use the terminologies even when negative exponents are present.}
Syntactically, polynomials form a strict subclass of the expressions definable using \eqref{equ:syntax probability prop.}, because the product of two polynomials is not a polynomial, but is a valid expression according to \eqref{equ:syntax probability prop.}.
A PSE $\varphi$ is \emph{division-free} if there is no division operator involved in $\varphi$.
The \emph{size} of an expression $\varphi$ is the total number of arithmatic operators (i.e.\ $+,-,\cdot,\div$) in $\varphi$.
We use $V_\varphi$ to denote the set of variables appearing in the expression $\varphi$, and for every $V\subseteq V_\varphi$ we define $\dom(V)\coloneqq \set{i\in \Q \mid \exists v_{ij}\in V \lor \exists v_{ki}\in V}$ as the set containing any state of the Markov chain that is involved in some variable in $V$. 

The semantics of a PSE $\varphi$ is interpreted \emph{statically} on the unknown Markov chain $M$:
we write $\varphi(M)$ to denote the evaluation or the value of $\varphi$ by substituting every variable $v_{ij}$ in $\varphi$ with $M_{ij}$.
E.g., for a Markov chain with state space $\set{1,2}$ and transition probabilities $M_{11}=0.2$, $M_{12}=0.8$, $M_{21}=0.4$, and $M_{22}=0.6$, 
the expression $\varphi = v_{11}-v_{21}$ has the evaluation $\varphi(M) = 0.2-0.4 = -0.2$.
We will assume that for every expression $(1\div \xi)$, $\xi(M)\neq 0$.

\smallskip
\noindent\textbf{Example: Group fairness.}
	Using PSEs, we can express the group fairness properties for the lending example described in Sec.~\ref{sec:motivating examples}, with the help of the Markov chain in the left subfigure of Fig.~\ref{fig:markov chain examples}:\\
	\smallskip
\begin{minipage}{\textwidth}
\renewcommand{\arraystretch}{1.5}
\begin{center}
	 \begin{tabular}{m{5cm} m{5cm}}
	 	\textbf{Disparate impact} \cite{feldman2015certifying}: & $v_{gy}\div v_{\overline{g}y}$\\
	 	\textbf{Demographic parity} \cite{dwork2012fairness}: & $v_{gy}-v_{\overline{g}y}$\\
	 \end{tabular}
	 \end{center}
\end{minipage}
	The equal opportunity criterion requires the following probability to be close to zero:
	$
	p = \pr(y\mid g,z)-\pr(y\mid \overline{g},z),
	$	
	which is tricky to monitor as $p$ contains the counter-factual probabilities representing ``the probability that an individual from a group would repay had the loan been granted."
	We apply Bayes' rule, and turn $p$ into the following equivalent form:
	$
	 p'=\frac{\pr(z\mid g,y)\cdot\pr(y\mid g)}{\pr(z\mid g)} - \frac{\pr(z\mid \overline{g},y)\cdot\pr(y\mid \overline{g})}{\pr(z\mid \overline{g})}.
	$
	Assuming $\pr(z\mid g)=c_1$ and $\pr(z\mid\overline{g})=c_2$, where $c_1$ and $c_2$ are known constants, the property $p'$ can be encoded as a PSE as below:\\
	\smallskip
\begin{minipage}{\textwidth}
\renewcommand{\arraystretch}{1.5}
	\begin{center}
	\begin{tabular}{m{5cm} m{5cm}}
			 	\textbf{Equal opportunity} \cite{hardt2016equality}: & $(v_{(gy)z}\cdot v_{gy})\div c_1 - ( v_{(\overline{g}y)z}\cdot v_{\overline{g}y}) \div c_2$.
	\end{tabular}
	\end{center}
\end{minipage}

\smallskip
\noindent\textbf{Example: Social burden.} 
	Using PSEs, we can express the social burden of the college admission example described in Sec.~\ref{sec:motivating examples}, with the help of the Markov chain depicted in the right subfigure of Fig.~\ref{fig:markov chain examples}:\\
	\begin{minipage}{\textwidth}
	\renewcommand{\arraystretch}{1.5}
	\begin{center}
		 \begin{tabular}{m{5cm} m{5cm}}
		 		\textbf{Social burden} \cite{milli2019social}: & $1\cdot v_{g1} + \ldots + N\cdot v_{gN}$.
		 \end{tabular}
	\end{center}
	\end{minipage}
	\renewcommand{\arraystretch}{1}


\subsection{The Monitoring Problem}\label{sec:problem}

Informally, our goal is to build monitors that observe a single long path of a Markov chain and, after each observation, output a new estimate for the value of the PSE.
Since the monitor's estimate is based on statistics collected from a finite path, the output may be incorrect with some probability, where the source of this probability is different between the frequentist and the Bayesian approaches.
In the frequentist approach, the underlying Markov chain is fixed (but unknown), and the randomness stems from the sampling of the observed path.
In the Bayesian approach, the observed path is fixed, and the randomness stems from the uncertainty about a prior specifying the Markov chain's parameters.
The commonality is that, in both cases, we want our monitors to estimate the value of the PSE up to an error with a fixed probabilistic confidence.

We formalize the monitoring problem separately for the two approaches.
A \emph{problem instance} is a triple $\tup{\Q,\varphi,\delta}$, where $\Q=\range{N}$ is a set of states, $\varphi$ is a PSE over $\Q$, and $\delta\in [0,1]$ is a constant.
In the frequentist approach, we use $\pr_s$ to denote the probability measure induced by \emph{sampling} of paths, and in the Bayesian approach we use $\pr_\theta$ to denote the probability measure induced by the \emph{prior} probability density function $p_\theta\colon \simplex(n-1)^n\to \mathbb{R}\cup \{\infty\}$ over the transition matrix of the Markov chain.
In both cases, the output alphabets of the monitors contain every real interval. 

\begin{prob}[Frequentist monitor]\label{prob:frequentist:quantitative}
	Suppose $\tup{\Q,\varphi,\delta}$ is a problem instance given as input.
	Design a monitor $\monitor$ such that for every Markov chain $\mc$ with transition probability matrix $\trm$ and for every finite path $\seq{x}\in \paths{\mc}$: 
	\begin{align}\label{equ:problem:frequentist:quantitative}
		\pr_{s,\monitor}\left(\varphi(M)\in \sem{\monitor}(\seq{x})\right) \geq 1-\delta, 
	\end{align}
	where $\pr_{s,\monitor}$ is the joint probability measure of $\pr_s$ and $\pr_\monitor$.
\end{prob}

\begin{prob}[Bayesian monitor]\label{prob:bayesian:quantitative}
	Suppose $\tup{\Q,\varphi,\delta}$ is a problem instance and $p_\theta$ is a prior density function, both given as inputs.
	Design a monitor $\monitor$ such that for every Markov chain $\mc$ with transition probability matrix $\trm$ and for every finite path $\seq{x}\in \paths{\mc}$:
	\begin{align}\label{equ:problem:bayesian:quantitative}
		\pr_{\theta,\monitor}\left(\varphi(M)\in \sem{\monitor}(\seq{x})\mid \seq{x}\right) \geq 1-\delta, 
	\end{align}
	where $\pr_{\theta,\monitor}$ is the joint probability measure of $\pr_\theta$ and $\pr_\monitor$.
\end{prob}

Notice that the state space of the Markov chain and the input alphabet of the monitor are the same, and so, many times, we refer to observed states as (input) symbols, and vice versa.
The estimate $[l,u]=\sem{\monitor}(\seq{x})$ is called the $(1-\delta)\cdot 100\%$ \emph{confidence interval} for $\varphi(M)$.\footnote{
While in the Bayesian setting \emph{credible intervals} would be more appropriate, we use confidence intervals due to uniformity and the relative ease of computation.
To relate the two, our confidence intervals are over-approximations of credible intervals (non-unique) that are centered around the posterior mean.}
The radius, given by $\varepsilon=0.5\cdot (u-l)$, is called the \emph{estimation error}, and the quantity $1-\delta$ is called the \emph{confidence}.
The estimate gets more precise as the error gets smaller and the confidence gets higher. 

In many situations, we are interested in a \emph{qualitative} question of the form ``is $\varphi(M)\leq c$?'' for some constant $c$.
We point out that, once the quantitative problem is solved, the qualitative questions can be answered using standard procedures by setting up a hypothesis test \cite[p.~380]{knight1999mathematical}.


\section{Frequentist Monitoring}\label{sec:frequentist:background}


Suppose the given PSE is only a single variable $\varphi=v_{ij}$, i.e., we are monitoring the probability of going from state $i$ to another state $j$.
The frequentist monitor $\monitor$ for $\varphi$ can be constructed in two steps: (1) empirically compute the average number of times the edge $(i,j)$ was taken per visit to the state $i$ on the observed path of the Markov chain, and (2) compute the $(1-\delta)\cdot 100\%$ confidence interval using statistical concentration inequalities.

\begin{wrapfigure}{r}{4.5cm}
	\vspace{-0.5cm}
	\includegraphics[scale=0.25]{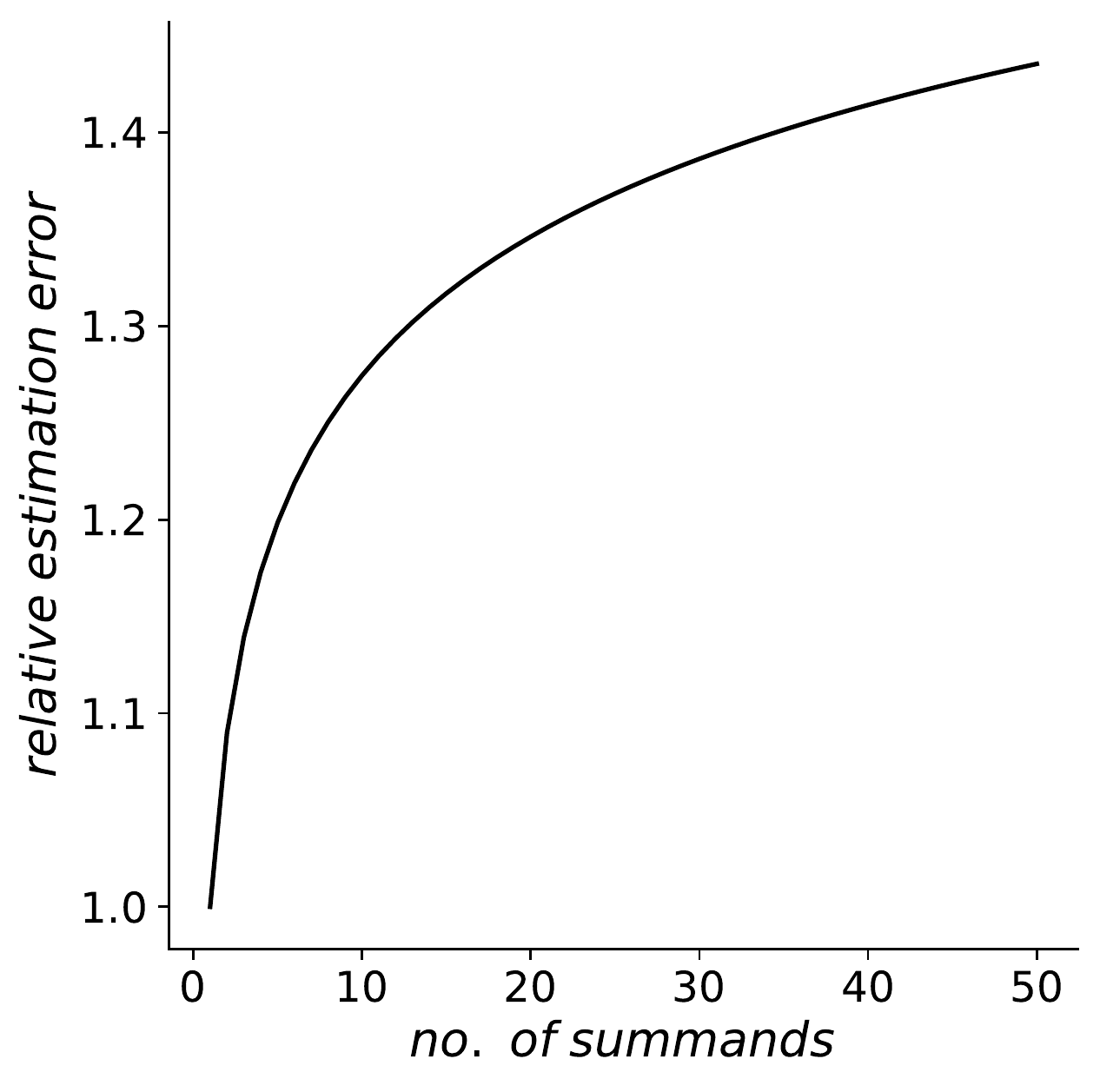}
	\caption{Variation of ratio of the est.\ error using the existing approach \cite{albarghouthi2019fairness} to est.\ error using our approach, w.r.t.\ the size of the chosen PSE.}
	\label{fig:frequentist comparison}
	\vspace{-0.5cm}
\end{wrapfigure}
Now consider a slightly more complex PSE $\varphi'=v_{ij}+v_{ik}$.
One approach to monitor $\varphi'$, proposed by Albarghouthi et al.~\cite{albarghouthi2019fairness}, would be to first compute the $(1-\delta)\cdot 100\%$ confidence intervals $[l_1,u_1]$ and $[l_2,u_2]$ separately for the two constituent variables $v_{ij}$ and $v_{ik}$, respectively.
Then, the $(1-2\delta)\cdot 100\%$ confidence interval for $\varphi'$ would be given by the sum of the two intervals $[l_1,u_1]$ and $[l_2,u_2]$, i.e., $[l_1+l_2,u_1+u_2]$; notice the drop in overall confidence due to the union bound.
The drop in the confidence level and the additional error introduced by the interval arithmetic accumulate quickly for larger PSEs, making the estimate unusable.
Furthermore, we lose all the advantages of having any dependence between the terms in the PSE.
For instance, by observing that $v_{ij}$ and $v_{ik}$ correspond to the mutually exclusive transitions $i$ to $j$ and $i$ to $k$, we know that $\varphi'(M)$ is always less than $1$, a feature that will be lost if we use plain merging of individual confidence intervals for $v_{ij}$ and $v_{ik}$.
We overcome these issues by estimating the value of the PSE as a whole as much as possible.
In Fig.~\ref{fig:frequentist comparison}, we demonstrate how the ratio between the estimation errors from the two approaches vary as the number of summands (i.e., $n$) in the PSE $\varphi=\sum_{i=1}^n v_{1n}$ changes; in both cases we fixed the overall $\delta$ to $0.05$ ($95\%$ confidence).
The ratio remains the same for different observation lengths.
Our approach is always at least as accurate as their approach \cite{albarghouthi2019fairness}, and is significantly better for larger PSEs.


\subsection{The Main Principle}


We first explain the idea for division-free PSEs, i.e., PSEs that do not involve any division operator; later we extend our approach to the general case.

\smallskip
\noindent\textbf{Divison-free PSEs:} 
In our algorithm, for every variable $v_{ij}\in V_\varphi$, we introduce a $\bernoulli{{M_{ij}}}$ random variable $Y^{ij}$ with the mean ${M_{ij}}$ unknown to us.
We make an observation $y^{ij}_p$ for every $p$-th visit to the state $i$ on a run, and if $j$ follows immediately afterwards then record $y^{ij}_p=1$ else record $y^{ij}_p=0$.
This gives us a sequence of observations $\seq{y}^{ij}=y^{ij}_1,y^{ij}_2,\ldots$ corresponding to the sequence of i.i.d.\ random variables $\seq{Y}^{ij}= Y^{ij}_1,Y^{ij}_2,\ldots$.
For instance, for the run $121123$ we obtain $\seq{y}^{12}=1,0,1$ for the variable $v_{12}$.

The heart of our algorithm is an aggregation procedure of every sequence of random variable $\set{\seq{Y}^{ij}}_{v_{ij}\in V_{\varphi}}$ to a single i.i.d.\ sequence $\seq{W}$ of an auxiliary random variable $W$, such that the mean of $W$ is $\mu_W =\expe(W) = \varphi(M)$.
We can then use known concentration inequalities on the sequence $\seq{W}$ to estimate $\mu_W$.
Since $\mu_{W}$ exactly equals $\varphi(M)$ by design, we obtain a tight concentration bound on $\varphi(M)$.
We informally explain the main idea of constructing $\seq{W}$ using simple examples; the details can be found in Alg.~\ref{alg:frequentist monitor}.

\smallskip
\noindent\textbf{Sum and difference:}
Let $\varphi = v_{ij} + v_{kl}$.
We simply combine $\seq{Y}^{ij}$ and $\seq{Y}^{kl}$ as $W_p = Y^{ij}_p + Y^{kl}_p$, so that $w_p = y^{ij}_p+y^{kl}_p$ is the corresponding observation of $W_p$.
Then $\mu_{W_p} = \varphi(M)$ holds, because $\mu_{W_p} = \expe(W_p) = \expe(Y^{ij}_p + Y^{kl}_p) = \expe(Y^{ij}_p) + \expe(Y^{kl}_p) = M_{ij} + M_{kl}$.
Similar approach works for $\varphi = v_{ij} - v_{kl}$.

\smallskip
\noindent\textbf{Multiplication:}
For multiplications, the same linearity principle will not always work, since for random variables $A$ and $B$, $\expe(A\cdot B)=\expe(A)\cdot \expe(B)$ \emph{only if} $A$ and $B$ are statistically independent, which will not be true for specifications of the form $\varphi = v_{ij}\cdot v_{ik}$.
In this case, the respective Bernoulli random variables $Y^{ij}_p$ and $Y^{ik}_p$ are dependent:
$\pr(Y^{ij}_p=1)\cdot \pr(Y^{ik}_p=1)=M_{ij}\cdot M_{ik}$, but $\pr(Y^{ij}_p=1 \land Y^{ik}_p=1)$ is always 0 (since \emph{both} $j$ and $k$ cannot be visited following the $p$-th visit to $i$).

To benefit from independence once again, we temporally shift one of the random variables by defining $W_p = Y^{ij}_{2p}\cdot Y^{ik}_{2p+1}$, with $w_p = y^{ij}_{2p}\cdot y^{ik}_{2p+1}$.
Since the random variables $Y^{ij}_{2p}$ and $Y^{ik}_{2p+1}$ are independent, as they use separate visits of state $i$, hence we obtain $\mu_{W_p} = M_{ij}\cdot M_{ik}$.
For independent multiplications of the form $\varphi = v_{ij}\cdot v_{kl}$ with $i\neq k$, we can simply use $W_p = Y^{ij}_{p}\cdot Y^{ik}_{p}$. 

In general, we use the ideas of aggregation and temporal shift on the syntax tree of the PSE $\varphi$, inductively. With an aggregated sequence of observations for the auxiliary variable $W$ for $\varphi$, we can find an estimate for $\varphi(M)$ using the Hoeffding's inequality.
We present the detailed algorithm of this monitor, namely \algfreqdivfree, in Alg.~\ref{alg:frequentist monitor division-free}. 

\smallskip
\noindent\textbf{The general case (PSEs with division operators):}
We observe that every arbitrary PSE $\varphi$ of size $n$ can be transformed into a semantically equivalent PSE of the form $\varphi_a+\frac{\varphi_b}{\varphi_c}$ of size $\mathcal{O}(n^22^n)$, where $\varphi_a$, $\varphi_b$, and $\varphi_c$ are all division-free. 
Once in this form, we can employ three different \algfreqdivfree monitors from Alg.~\ref{alg:frequentist monitor division-free} to obtain separate interval estimates for $\varphi_a$, $\varphi_b$, and $\varphi_c$, which are then combined using standard interval arithmetic and the resulting confidence of the estimate is obtained through the union bound.
The steps for constructing the (general-case) \texttt{FrequentistMonitor} are shown in Alg.~\ref{alg:frequentist monitor}, and the detailed analysis can be found in the proof of Thm.~\ref{thm:frequentist:soundness}.

\smallskip
\noindent\textbf{Bounding memory:} 
Consider a PSE $\varphi=v_{ij}+v_{kl}$.
The outcome $w_p$ for $\varphi$ can only be computed when both the Bernoulli outcomes $y^{ij}_p$ and $y^{kl}_p$ are available.
If at any point only one of the two is available, then we need to store the available one so that it can be used later when the other one gets available.
It can be shown that the storage of ``unmatched'' outcomes may need unbounded memory.

To bound the memory, we use the insight that a \emph{random reshuffling} of the i.i.d.\ sequence $y^{ij}_p$ would still be i.i.d.\ with the same distribution, so that we do not need to store the exact order in which the outcomes appeared.
Instead, for every $v_{ij}\in V_\varphi$, we only store the number of times we have seen the state $i$ and the edge $(i,j)$ in counters $c_i$ and $c_{ij}$, respectively.
Observe that $c_i\geq\sum_{v_{ik}\in V_\varphi} c_{ik}$, where the possible difference accounts for the visits to irrelevant states, denoted as a dummy state $\top$.
Given $\set{c_{ik}}_{k}$, whenever needed, we generate in $x_i$ a \emph{random reshuffling} of the sequence of states, together with $\top$, seen after the past visits to $i$. 
From the sequence stored in $x_i$, for every $v_{ik}\in V_\varphi$, we can consistently determine the value of $y^{ik}_p$ (consistency dictates $y^{ik}_p=1\Rightarrow y^{ij}_p=0$).
Moreover, we reuse space by resetting $x_i$ whenever the sequence stored in $x_i$ is no longer needed.
\ifarxiv
	It is shown in the proof of Thm.~\ref{thm:frequentist:complexity} in App.~\ref{app:freq:proofs} that the size of every $x_i$ can be at most the size of the expression.
\else
	It can be shown that the size of every $x_i$ can be at most the size of the expression \cite[Proof of Thm.~2]{??}.
\fi
This random reshuffling of the observation sequences is the cause of the probabilistic transitions of the frequenitst monitor.


\subsection{Implementation of the Frequentist Monitor}

Fix a problem instance $\tup{\Q,\varphi,\delta}$, with size of $\varphi$ being $n$. 
Let $\varphi$ be transformed into $\varphi^l$ by relabeling duplicate occurrences of $v_{ij}$ using distinct labels $v_{ij}^1,v_{ij}^2,\ldots$.
The set of labeled variables in $\varphi^l$ is $V_\varphi^l$, and $|V_\varphi^l|=\mathcal{O}(n)$.
Let $\mathit{SubExpr}(\varphi)$ denote the set of every subexpression in the expression $\varphi$, and use $[l_\varphi, u_\varphi]$ to denote the range of values the expression $\varphi$ can take for every valuation of every variable as per the domain $[0,1]$.
Let $\mathit{Dep}(\varphi)=\set{i\mid \exists v_{ij}\in V_{\varphi}}$, and every subexpression $\varphi_1\cdot\varphi_2$ with $\mathit{Dep}(\varphi_1)\cap \mathit{Dep}(\varphi_2)\neq\emptyset$ is called a \emph{dependent multiplication}.

Implementation of $\algfreqdivfree$ in Alg.~\ref{alg:frequentist monitor division-free} has two main functions. $\mathit{Init}$ initializes the registers. $\mathit{Next}$ implements the transition function of the monitor, which attempts to compute a new observation $w$ for $\seq{W}$ (Line~\ref{alg:freq:next:eval}) after observing a new input $\sigma'$, and if successful it updates the output of the monitor by invoking the $\mathit{UpdateEst}$ function.
In addition to the registers in $\mathit{Init}$ and $\mathit{Next}$ labeled in the pseudocode, following registers are used internally:
\begin{itemize}[noitemsep,topsep=0pt,parsep=0pt,partopsep=0pt]
	\item $x_i,\, i\in \mathit{Dom}(V_{\varphi})$: reshuffled sequence of states that followed $i$.
	\item $t^{l}_{ij}$: the index of $x_i$ that was used to obtain the latest outcome of $v_{ij}^l$.
\end{itemize}


\newcommand{\ans}{\ensuremath{\mathit{ans}}}
\newcommand{\eval}{\ensuremath{\mathit{Eval}}}
\newcommand{\extractoutcome}{\ensuremath{\mathit{ExtractOutcome}}}
\newcommand{\timescale}{\ensuremath{\mathbb{T}}}
\newcommand{\true}{\ensuremath{\mathit{true}}}
\newcommand{\false}{\ensuremath{\mathit{false}}}
\renewcommand{\algorithmicensure}{\textbf{Output:}}
	\begin{algorithm}
	\caption{\algfreqdivfree}
	\label{alg:frequentist monitor division-free}
	\begin{minipage}{0.4\textwidth}
		\begin{algorithmic}[1] 
			\renewcommand{\algorithmicrequire}{\textbf{Parameters:}}
			\Require $\Q,\varphi,\delta$
			\Ensure $\verdict$
			\Function{$\mathit{Init}(\sigma)$}{}
			\State $\varphi^l\xleftarrow{\text{unique labeling}} \varphi$
			\ForAll{$v_{ij}\in V_{\varphi}$}
				\State $c_{ij}\gets 0$ \Comment{$\#$ of $(i,j)$}
				\State $c_{i}\gets 0$ \Comment{$\#$ of $i$}
			\EndFor
			\State $n\gets 0$ \Comment{length of $\seq{w}$}
			\State $\sigma \gets \sigma$ \Comment{prev. symbol}
			\State $\mu_\verdict \gets \bot$ \Comment{est. mean}
			\State $\varepsilon_\verdict\gets\bot$ \Comment{est. error}
			\State $\mathit{ResetX}()$ \Comment{reset $x_i$-s}
			\State Compute $l_\varphi,u_\varphi$ \Comment{int.\ arith.}
			\EndFunction
		\end{algorithmic}
	\end{minipage}
	\begin{minipage}{0.6\textwidth}
		\begin{algorithmic}[1] 
			\Function{$\mathit{Next}$}{$\sigma'$}
					\State $c_{\sigma}\gets c_{\sigma}+1$ \Comment{update counters} \label{alg:freq:next:a}
					\State $c_{\sigma\sigma'}\gets c_{\sigma\sigma'}+1$ \label{alg:freq:next:b}
			\State $w\gets \mathit{Eval}(\varphi^l)$ \label{alg:freq:next:eval}
			\If{$w \neq \bot$} 
				\State $n\gets n+1$
				\State $\verdict\gets\mathit{UpdateEst}(w,n)$ \label{alg:freq:next:l}
				\State $\mathit{ResetX}()$
			\EndIf
			\State $\sigma \gets \sigma'$
			\State \Return $\verdict$
			\EndFunction
		\end{algorithmic}
	\end{minipage}
			\rule{\textwidth}{0.4pt}

	\begin{minipage}{0.57\textwidth}
		\begin{algorithmic}[1]
			\Function{$\mathit{Eval}$}{$\varphi^l$}
				\If{$r_{\varphi^l} = \bot$}
					\If{$\varphi^l\equiv\varphi^l_1+\varphi^l_2$}
						 \State $r_{\varphi^l}\gets\mathit{Eval}(\varphi^l_1)+\mathit{Eval}(\varphi^l_2)$\label{alg:eval:+}
					\ElsIf{$\varphi^l\equiv\varphi^l_1-\varphi^l_2$}
						 \State $r_{\varphi^l}\gets\mathit{Eval}(\varphi^l_1)-\mathit{Eval}(\varphi^l_2)$ \label{alg:eval:-}
					\ElsIf{$\varphi^l\equiv\varphi^l_1\cdot \varphi^l_2$}
						\If{$\depends(V_{\varphi_1}^l)\cap \depends(V_{\varphi_2}^l)=\emptyset$} 
							 \State $r_{\varphi^l}\gets\mathit{Eval}(\varphi^l_1)\cdot \mathit{Eval}(\varphi^l_2)$\label{alg:eval:ind *}
						\Else  \Comment{dep. mult.}
							\For{$v_{ij}^l\in V_{\varphi_2}^l\cap \depends(V_{\varphi_1}^l)$}\label{alg:eval:dep * start}
								\State $t_{ij}^l\gets \max(\set{t_{ik}^m \mid v_{ik}^m\in V_{\varphi_1}^l})$
								\State $t_{ij}^l\gets t_{ij}^l + 1$ \Comment{make indep.} 
							\EndFor
							\State $r_{\varphi^l}\gets \mathit{Eval}(\varphi^l_1)\cdot \mathit{Eval}(\varphi^l_2)$ \label{alg:eval:dep *}
						\EndIf
				\ElsIf{$\varphi^l\equiv v_{ij}^l$}
					\If{$x_{i}[t_{ij}^l+1] = \bot$}
						\State $\mathit{ExtractOutcome}(x_i,t_{ij}^l+1)$ \label{alg:eval:extract}
					\EndIf
					\If{$x_{i}[t_{ij}^l+1]=j\neq \bot$}\label{alg:eval:var}
						\State $r_{\varphi^l}\gets 1$
					\Else 
						\State $r_{\varphi^l}\gets 0$
					\EndIf
				\ElsIf{$\varphi^l\equiv c$}
					\State $r_{\varphi^l}\gets c$ \label{alg:eval:const}
				\EndIf
				\EndIf
				\State \Return $r_{\varphi^l}$
			\EndFunction
		\end{algorithmic}
	\end{minipage}
	\begin{minipage}{0.42\textwidth}
		\begin{algorithmic}[1]
			\vspace{0.3cm}
			\Function{$\mathit{UpdateEst}$}{$w,n$} 
				\State $\mu_\verdict \gets \frac{\mu_\verdict\cdot (n-1)+w}{n}$ \label{alg:freq:output:update mu}
					\State $\varepsilon_\verdict\gets \sqrt{-\frac{(u_\varphi-l_\varphi)^2}{2n}\cdot \ln\left(\frac{\delta}{2}\right)}$ \label{alg:freq:output:hoeffding}
				\State \Return $[\mu_\verdict\pm\varepsilon_\verdict]$
			\EndFunction
		\end{algorithmic}
		\vspace{0.2cm}
		\begin{algorithmic}[1]
			\Function{$\mathit{ExtractOutcome}$}{$x_i,t$} \Comment{generate a shuffled sequence of symbols seen after $i$ so that $|x_i|=t$}
				\State Let $U\gets \set{j\in \Q \mid v_{ij}\in V_\varphi}$
				\For{$p=|x_{i}|+1,\ldots,t$}
					\State $q\gets $ \parbox[t]{0.5\linewidth} {
									$ \forall u\in U\;.\;$\\
									 $\text{pick } u \text{ w/ prob.\ } \frac{c_{iu}}{c_i}, $\\ 
									$ \text{pick }\top \text{ w/ prob.\ }\frac{(c_i-\sum_j c_{ij})}{c_i}$} \label{alg:extractoutcome:rand}
					\State $c_i\gets c_i-1$
					\If{$q\neq \top$}
						\State $c_{iq}\gets c_{iq}-1$
					\EndIf
					\State $x_i[|x_i|+1]\gets q$
				\EndFor
			\EndFunction
		\end{algorithmic}
		\vspace{0.2cm}
		\begin{algorithmic}[1]
			\Function{$\mathit{ResetX}()$}{}
				\ForAll{$i\in \dom(V_{\varphi})$}
					\State $x_{i}\gets \emptyset$				
				\EndFor
				\ForAll{$v_{ij}^l\in V_\varphi^l$}
					\State $t_{ij}^l\gets 0$
				\EndFor
			\EndFunction
		\end{algorithmic}
	\end{minipage}\\
%
\end{algorithm}

\begin{algorithm}
	\caption{\texttt{FrequentistMonitor}}
	\label{alg:frequentist monitor}
	\begin{minipage}{0.54\textwidth}
		\begin{algorithmic}[1] 
			\renewcommand{\algorithmicrequire}{\textbf{Parameters:}}
			\Require $\Q,\varphi,\delta$ 
			\Ensure $\verdict$
			\Function{$\mathit{Init}(\sigma)$}{}
				\State $\varphi_a + \frac{\varphi_b}{\varphi_c}\xleftarrow{\text{change form}} \varphi^l\xleftarrow{\text{labeling}} \varphi$
				\State $\monitor_a \gets \algfreqdivfree(\Q,\varphi_a,\delta/3)$
				\State $\monitor_b \gets \algfreqdivfree(\Q,\varphi_b,\delta/3)$ \label{line:alg:freq:phi_c}
				\State $\monitor_c \gets \algfreqdivfree(\Q,\varphi_c,\delta/3)$
				\State $\monitor_a.\mathit{Init}(\sigma)$
				\State $\monitor_b.\mathit{Init}(\sigma)$
				\State $\monitor_c.\mathit{Init}(\sigma)$
			\EndFunction
		\end{algorithmic}
	\end{minipage}
	\begin{minipage}{0.48\textwidth}
		\begin{algorithmic}[1]
			\Function{$\mathit{Next}$}{$\sigma'$}
				\State $[\mu_a\pm \varepsilon_a]\gets \monitor_a.\mathit{Next}(\sigma')$
				\State $[\mu_b\pm \varepsilon_b]\gets \monitor_b.\mathit{Next}(\sigma')$
				\State $[\mu_c\pm \varepsilon_c]\gets \monitor_c.\mathit{Next}(\sigma')$
				\If{$\mu_a\neq \bot \wedge \mu_b\neq \bot \wedge \mu_c\neq \bot$}
					\State $[\mu_\verdict\pm\varepsilon_\verdict] \gets [\mu_a\pm \varepsilon_a] + \frac{[\mu_b\pm \varepsilon_b]}{[\mu_c\pm \varepsilon_c]}$
				\EndIf
				\State \Return $[\mu_\verdict\pm\varepsilon_\verdict]$
			\EndFunction
		\end{algorithmic}
	\end{minipage}
\end{algorithm}

\noindent
Now, we summarize the main results for the frequentist monitor.

\begin{theorem}[Correctness]\label{thm:frequentist:soundness}
	Let $(\Q,\varphi,\delta)$ be a problem instance.
	Alg.~\ref{alg:frequentist monitor} implements a monitor for $(\Q,\varphi,\delta)$ that solves Prob.~\ref{prob:frequentist:quantitative}.
\end{theorem}

\begin{theorem}[Computational resources]\label{thm:frequentist:complexity}
	Let $(\Q,\varphi,\delta)$ be a problem instance and $\monitor$ be the monitor implemented using the \emph{\texttt{FrequentistMonitor}} routine of Alg.~\ref{alg:frequentist monitor}.
	Suppose the size of $\varphi$ is $n$.
	The monitor $\monitor$ requires $\mathcal{O}(n^42^{2n})$ registers, and takes $\mathcal{O}(n^42^{2n})$ time to update its output after receiving a new input symbol.
	For the special case of $\varphi$ containing at most one division operator (division by constant does not count), $\monitor$ requires only $\mathcal{O}(n^2)$ registers, and takes only $\mathcal{O}(n^2)$ time to update its output after receiving a new input symbol.
\end{theorem}

There is a tradeoff between the estimation error, the confidence, and the length of the observed sequence of input symbols.
For instance, for a fixed confidence, the longer the observed sequence is, the smaller is the estimation error.
The following theorem establishes a lower bound on the length of the sequence for a given upper bound on the estimation error and a fixed confidence. 

\begin{theorem}[Convergence speed]\label{thm:frequentist:convergence}
	Let $(\Q,\varphi,\delta)$ be a problem instance where $\varphi$ does not contain any division operator, and let $\monitor$ be the monitor computed using Alg.~\ref{alg:frequentist monitor}.
	Suppose the size of $\varphi$ is $n$.
	For a given upper bound on estimation error $\overline{\varepsilon}\in \mathbb{R}$,
	the minimum number of visits to every state in $\dom(V_{\varphi})$ for obtaining an output with error at most $\overline{\varepsilon}$ and confidence at least $1-\delta$ on any path is given by:
	\begin{equation}	
		 -\frac{(u_\varphi-l_\varphi)^2\ln\left(\frac{\delta}{2}\right)n}{2\overline{\varepsilon}^2},
	\end{equation}
	where $[l_\varphi, u_\varphi]$ is the set of possible values of $\varphi$ for every valuation of every variable (having domain $[0,1]$) in $\varphi$.
\end{theorem}

The bound follows from the Hoeffding's inequality, together with the fact that every dependent multiplication increments the required number of samples by $1$.
A similar bound for the general case with division is left open.


\section{Bayesian Monitoring}
\label{sec:bayesian:background}


Fix a problem instance $\tup{\Q=\range{N},\varphi,\delta}$.
Let $\mcset=\simplex(N-1)^{N}$ be the shorthand notation for the set of transition probability matrices of the Markov chains with state space $\Q$.
Let $p_\theta\colon\mcset\to [0,1]$ be the prior probability density function over $\mcset$, which is assumed to be specified using the matrix beta distribution 
\ifarxiv
	(the definition is standard \cite[pp.~280]{insua2012bayesian} and has been included in App.~\ref{app:matrix beta} for completeness).
\else
	(the definition can be found in standard textbooks on Bayesian statistics \cite[pp.~280]{insua2012bayesian}).
\fi
Let $\mathbb{1}$ be a matrix, with its size dependent on the context, whose every element is $1$.
We make the following common assumption \cite[p.~50]{gomez2014b,insua2012bayesian}:

\begin{assumption}[Prior]\label{assump:bayes:matrix beta}
	We are given a parameter matrix $\theta\geq \mathbb{1}$, and $p_\theta$ is specified using the matrix beta distribution with parameter $\theta$.
	Moreover, the initial state of the Markov chain is fixed.
\end{assumption}

When $\theta=\mathbb{1}$, then $p_\theta$ is the uniform density function over $\mcset$.
After observing a path $\seq{x}$, using Bayes' rule we obtain the \emph{posterior} density function $p_\theta(\cdot\mid \seq{x})$, which is known to be efficiently computable due to the so-called conjugacy property that holds due to  
\ifarxiv
	Assump.~\ref{assump:bayes:matrix beta} (see App.~\ref{app:bayes:inference} for details). 
\else
	Assump.~\ref{assump:bayes:matrix beta}.
\fi
From the posterior density, we obtain the expected posterior semantic value of $\varphi$ as:
$\expe_{\theta}(\varphi(\trm)\mid \seq{x}) \coloneqq \int_{\mcset} \varphi(\trm)\cdot\prd(M\mid\seq{x})dM$.
The heart of our Bayesian monitor is an efficient incremental computation of $\expe_\theta(\varphi(M)\mid \seq{x})$---free from numerical integration.
Once we can compute $\expe_\theta(\varphi(M)\mid \seq{x})$, we can also compute the posterior variance $S^2$ of $\varphi(M)$ using the known expression $S^2=\expe_\theta(\varphi^2(M)\mid\seq{x})-\expe_\theta(\varphi(M)\mid\seq{x})$, which enables us to compute a confidence interval for $\varphi(M)$ using the Chebyshev's 
\ifarxiv
	inequality (see Prop.~\ref{prop:bayes:conf interval using chebyshev} in App.~\ref{app:bayes:soundness}).
\else
	inequality.
\fi
In the following, we summarize our procedure for estimating $\expe_\theta(\varphi(M)\mid \seq{x})$.


\subsection{The Main Principle}
\label{sec:bayesian:principle}

The incremental computation of $\expe_\theta(\varphi(M)\mid \seq{x})$ is implemented in $\mathtt{BayesExpMonitor}$. 
We first transform the expression $\varphi$ into the polynomial form $\varphi'=\sum_l \kappa_l\xi_l$, where $\set{\kappa_l}_l$ are the weights and $\set{\xi_l}_l$ are monomials.
If the size of $\varphi$ is $n$ then the size of $\varphi'$ is $\mathcal{O}(n2^{\frac{n}{2}})$.
Then we can use linearity to compute the overall expectation as the weighted sum of expectations of the individual monomials: $\expe_\theta(\varphi(\trm)\mid\seq{x}) = \expe_\theta(\varphi'(\trm)\mid \seq{x})=\sum_l \kappa_l\expe_\theta(\xi_l(M)\mid\seq{x})$. 
In the following, we summarize the procedure for estimating $\expe_\theta(\xi(M)\mid\seq{x})$ for every monomial $\xi$.

Let $\xi$ be a monomial, and let $\seq{x}ab\in \Q^*$ be a sequence of states.
We use $d_{ij}$ to store the exponent of the variable $v_{ij}$ in the monomial $\xi$, and define $d_a\coloneqq \sum_{j\in \range{N}} d_{aj}$.
Also, we record the sets of $(i,j)$-s and $i$-s with positive and negative $d_{ij}$ and $d_{i}$ entries:
$D_i^+\coloneqq \set{j\mid d_{ij}>0}$, $D_i^-\coloneqq \set{j\mid d_{ij}<0}$, $D^+\coloneqq \set{i\mid d_i>0}$, and $D^-\coloneqq \set{i\mid d_i<0}$.

For any given word $\seq{w}\in \Q^*$, let $c_{ij}({\seq{w}})$ denote the number of $ij$-s in $\seq{w}$ and let $c_{i}({\seq{w}})\coloneqq \sum_{j\in\Q} c_{ij}({\seq{w}})$.
Define $\overline{c}_i({\seq{w}})\coloneqq c_i({\seq{w}}) + \sum_{j\in\range{N}}\theta_{ij}$ and $\overline{c}_{ij}({\seq{w}})\coloneqq c_{ij}({\seq{w}})+\theta_{ij}$.
Let $\uf\colon \Q^*\to \mathbb{R}$ be defined as:
\begin{align}\label{equ:bayes:def H}
	\uf(\seq{w}) \coloneqq 
	\frac{
		\prod_{i= 1}^N \prod_{j\in D_i^+} \Myperm[(\overline{c}_{ij}({\seq{w}})-1)+|d_{ij}|]{|d_{ij}|}
		}
		{
		\prod_{i\in D^+} \Myperm[(\overline{c}_i({\seq{w}})-1)+|d_i|]{|d_i|}
		}
	\cdot \frac{
		\prod_{i\in D^-}\Myperm[(\overline{c}_i({\seq{w}})-1)]{|d_i|}
		}
		{
		\prod_{i= 1}^N\prod_{j\in D_i^-}\Myperm[(\overline{c}_{ij}({\seq{w}})-1)]{|d_{ij}|}
		},
\end{align}
where $\Myperm[n]{k}\coloneqq \frac{n!}{(n-k)!}$ is the number of permutations of $k>0$ items from $n>0$ objects, for $k\leq n$, and we use the convention that for $S=\emptyset$, $\prod_{s\in S} \ldots = 1$. 
Below, in Lem.~\ref{thm:bayes:incremental computation}, we establish that $\expe_\theta(\xi(\trm)\mid\seq{w}) = \uf(\seq{w})$, and present an efficient incremental scheme to compute $\expe_\theta(\xi(\trm)\mid\seq{x}ab)$ from $\expe_\theta(\xi(\trm)\mid\seq{x}a)$.


\begin{lemma}[Incremental computation of $\expe(\cdot\mid \cdot)$]\label{thm:bayes:incremental computation}
	If the following consistency condition
	\setlength{\abovedisplayskip}{0pt}
\setlength{\belowdisplayskip}{0pt}
\begin{align}\label{equ:bayesian:consistency condition}
	\forall i,j\in \range{N}\;.\; \overline{c}_{ij}({\seq{w}}) + d_{ij} > 0
\end{align}
	is met, then the following holds:
\begin{align}\label{equ:bayesian:recurrence equation}
	\expe(\xi(M)\mid \seq{x}ab)=
	\uf(\seq{x}ab) = 
	\uf(\seq{x}a)\cdot \frac{\overline{c}_{ab}({\seq{x}})+d_{ab}}{\overline{c}_{ab}({\seq{x}})} \cdot \frac{\overline{c}_a({\seq{x}})}{\overline{c}_a({\seq{x}})+d_a}.
\end{align}
\end{lemma}

Cond.~\eqref{equ:bayesian:consistency condition} guarantees that the permutations in \eqref{equ:bayes:def H} are well-defined.
The first equality in \eqref{equ:bayesian:recurrence equation} follows from Marchal et al.~\cite{marchal2017sub}, and the rest uses the conjugacy of the 
\ifarxiv
	prior (proof in App.~\ref{app:bayes:h & update}).
\else
	prior.
\fi
Lem.~\ref{thm:bayes:incremental computation} forms the basis of the efficient update of our Bayesian monitor.
Observe that on any given path, once \eqref{equ:bayesian:consistency condition} holds, it continues to hold forever.
Thus, initially the monitor keeps updating $\uf$ internally without outputting anything.
Once \eqref{equ:bayesian:consistency condition} holds, it keeps outputting $\uf$ from then on.


\begin{algorithm}[t]
	\caption{$\mathtt{BayesExpMonitor}$}
	\label{alg:bayesian expectation monitor}
	\begin{minipage}{0.45\textwidth}
		\begin{algorithmic}[1] 
			\renewcommand{\algorithmicrequire}{\textbf{Parameters:}}
			\Require $\Q,\varphi = \sum_{l=1}^p \kappa_l\xi_l,\theta$ 
			\Ensure $E$
			\Function{$\mathit{Init}(\sigma=1)$}{}
			\For{$v_{ij}\in V_{\varphi}$}
				\State $\overline{c}_{ij}\gets \theta_{ij}$
				\State $\overline{c}_{i}\gets \sum_{j\in \range{N}}\theta_{ij}$
				\State $m_{ij}\gets \min_{l\in \range{p}}d_{ij}^l$ \Comment{cache} \label{alg:bayes:exp monitor:init:m}
			\EndFor
			\State $\activ\gets \false$ \Comment{eq.~\ref{equ:bayesian:consistency condition} not true}
			\State $\sigma \gets \sigma$ \Comment{prev.\ state}
			\State $E \gets \bot$ \Comment{expect.\ val.}
			\EndFunction
		\end{algorithmic}
	\end{minipage}
	\hfill
	\begin{minipage}{0.55\textwidth}
		\begin{algorithmic}[1] 
			\Function{$\mathit{Next}$}{$\sigma'$}
			\State $\overline{c}_{\sigma}\gets \overline{c}_{\sigma}+1$ \Comment{update counters}
			\State $\overline{c}_{\sigma\sigma'}\gets \overline{c}_{\sigma\sigma'}+1$ 
			\If{$\activ=\false$}
				\If{$\left(\forall v_{ij}\in V_{\varphi}\;.\;\overline{c}_{ij}+m_{ij}>0\right)$} \label{alg:bayes:exp monitor:next:check consistency}
					\State $\activ\gets \true$ \Comment{eq.~\ref{equ:bayesian:consistency condition} is true}
                    \For{$l\in \range{p}$} \Comment{eq.~\ref{equ:bayes:def H}}
				        \State {\scriptsize$h^l\gets \uf^l\left(\set{\overline{c}_{ij}}_{i,j},\set{\overline{c}_i}_{i} \right)$}
			        \EndFor
				\EndIf
			\Else
					\For{$l\in\range{p}$} \Comment{eq.~\ref{equ:bayesian:recurrence equation}}
						\State $h^l\gets h^l\cdot\frac{\overline{c}_{\sigma\sigma'}-1+d_{\sigma\sigma'}^l}{\overline{c}_{\sigma\sigma'}-1} \cdot \frac{\overline{c}_\sigma-1}{\overline{c}_\sigma-1+d_\sigma^l}$
					\EndFor
		
			\EndIf
			\If{$\activ=\true$}
				\State $E\gets \sum_{l=1}^p \kappa_l\cdot h^l$ \Comment{overall expect.}
			\EndIf
			\State $\sigma \gets \sigma'$
			\State \Return $E$
			\EndFunction
		\end{algorithmic}
	\end{minipage}
\end{algorithm}


\subsection{Implementation of the Bayesian Monitor}

We present the Bayesian monitor implementation in $\mathtt{BayesConfIntMonitor}$ (Alg.~\ref{alg:bayesian monitor}), which invokes $\mathtt{BayesExpMonitor}$ (Alg.~\ref{alg:bayesian expectation monitor}) as subroutine. 
$\mathtt{BayesExpMonitor}$ computes the expected semantic value of an expression $\varphi$ in polynomial form, by computing the individual expected value of each monomial using Prop.~\ref{thm:bayes:incremental computation}, and combining them using the linearity property.
We drop the arguments from $\overline{c}_i(\cdot)$ and $\overline{c}_{ij}(\cdot)$ and simply write $\overline{c}_i$ and $\overline{c}_{ij}$ as constants associated to appropriate words.
The symbol $m_{ij}$ in Line~\ref{alg:bayes:exp monitor:init:m} of $\mathit{Init}$ is used as a bookkeeping variable for quickly checking the consistency condition (Eq.~\ref{equ:bayesian:consistency condition}) in Line~\ref{alg:bayes:exp monitor:next:check consistency} of $\mathit{Next}$.
In $\mathtt{BayesConfIntMonitor}$, we compute the expected value and the variance of $\varphi$, by invoking $\mathtt{BayesExpMonitor}$ on $\varphi$ and $\varphi^2 $ respectively, and then compute the confidence interval using the Chebyshev's  
\ifarxiv
	inequality (Thm.~\ref{thm:chebyshev with true variance} in App.~\ref{app:confidence intervals}).
\else
	inequality.
\fi
It can be observed in the $\mathit{Next}$ subroutines of $\mathtt{BayesConfIntMonitor}$ and $\mathtt{BayesExpMonitor}$ that a deterministic transition function suffices for the Bayesian monitors.

\begin{algorithm}[t]
	\caption{$\mathtt{BayesConfIntMonitor}$}
	\label{alg:bayesian monitor}
	\begin{minipage}{0.55\textwidth}
		\begin{algorithmic}[1]
			\renewcommand{\algorithmicrequire}{\textbf{Parameters:}}
			\Require $\Q,\varphi,\theta$
			\Ensure $\verdict$
			\Function{$\mathit{Init}(\sigma=1)$}{}
				\State $\overline{\varphi}\xleftarrow[]{\text{polyn.}}\varphi$, $\overline{\varphi^2}\xleftarrow[]{\text{polyn.}}\varphi^2$ \Comment{polyn.\ form}
				\State $\mathit{EXP}\gets \mathtt{BayesExpMonitor}(\Q,\overline{\varphi},\theta)$
				\State $\mathit{EXP2}\gets \mathtt{BayesExpMonitor}(\Q,\overline{\varphi^2},\theta)$
				\State $\mathit{EXP}.\mathit{Init}(\sigma)$
				\State $\mathit{EXP2}.\mathit{Init}(\sigma)$
				\State $\verdict\gets \bot$
			\EndFunction
		\end{algorithmic}
	\end{minipage}
	\hfill
	\begin{minipage}{0.45\textwidth}
		\begin{algorithmic}[1]
			\Function{$\mathit{Next}$}{$\sigma'$}
				\State $E\gets \mathit{EXP}.\mathit{Next}(\sigma')$
				\State $E2\gets \mathit{EXP2}.\mathit{Next}(\sigma')$
				\If{$E\neq \bot$ and $E2\neq \bot$}
					\State $S\gets E2 - E^2$ \Comment{variance}
					\State $\verdict\gets \left[ E\pm \sqrt{\frac{S}{\delta}} \right]$ \Comment{Chebysh.}
				\EndIf
				\State \Return $\verdict$
			\EndFunction
		\end{algorithmic}
	\end{minipage}
\end{algorithm}

\begin{theorem}[Correctness]\label{thm:bayesian:soundness}
	Let $(\Q,\varphi,\delta)$ be a problem instance, and $p_\theta$ be given as the prior distribution which satisfies Assump.~\ref{assump:bayes:matrix beta}.
	Alg.~\ref{alg:bayesian monitor} produces a monitor for $(\Q,\varphi,\delta)$ that solves Prob.~\ref{prob:bayesian:quantitative}.
\end{theorem}

\begin{theorem}[Computational resources]\label{thm:bayesian:complexity}
	Let $(\Q,\varphi,\delta)$ be a problem instance and $\monitor$ be the monitor computed using the $\mathtt{BayesConfIntMonitor}$ routine of Alg.~\ref{alg:bayesian monitor}.
	Suppose the size of $\varphi$ is $n$.
	The monitor $\monitor$ requires $\mathcal{O}(n^22^n)$ registers, and takes $\mathcal{O}(n^22^n)$ time to update its output after receiving a new input symbol.
	For the special case of $\varphi$ being in polynomial form, $\monitor$ requires only $\mathcal{O}(n^2)$ registers, and takes only $\mathcal{O}(n^2)$ time to update its output after receiving a new input symbol.
\end{theorem}

A bound on the convergence speed of the Bayesian monitor is left open.
This would require a bound on the change in variance with respect to the length of the observed path, which is not known for the general case of PSEs.
Note that the efficient (quadratic) cases are different for the frequentist and Bayesian monitors, 
suggesting the use of different monitors for different specifications.


\section{Experiments}
\label{sec:experiments}

We implemented our frequentist and Bayesian monitors in a tool written in Rust, and used the tool to design monitors for the lending and the college admission examples taken from the literature \cite{liu2018delayed,milli2019social} (described in Sec.~\ref{sec:motivating examples}).
The generators are modeled as Markov chains (see Fig.~\ref{fig:markov chain examples})---unknown to the monitors---capturing the sequential interactions between the decision-makers (i.e., the bank or the college) and their respective environments (i.e., the loan applicants or the students), as described by D'Amour et al.~\cite{damour2020fairness}. 
The setup of the experiments is as follows:
We created a multi-threaded wrapper program, where one thread simulates one long run of the Markov chain, and a different thread executes the monitor.
Every time a new state is visited by the Markov chain on the first thread, the information gets transmitted to the monitor on the second thread, which then updates the output.
The experiments were run on a Macbook Pro 2017 equipped with a 2,3 GHz Dual-Core Intel Core i5 processor and 8GB RAM.

\begin{figure}[t]
	\caption{
	The plots show the $95\%$ confidence intervals estimated by the monitors over time, averaged over $10$ different sample paths, for the lending with demographic parity (left), lending with equalized opportunity (middle), and the college admission with social burden (right) problems.	
	The horizontal dotted lines are the ground truth values of the properties, obtained by analyzing the Markov chains used to model the systems (unknown to the monitors). 
	The table summarizes various performance metrics.
	\label{fig:experiments}
	}
	\begin{minipage}[t]{\textwidth}
		\vspace{0pt}
		\includegraphics[scale=0.275]{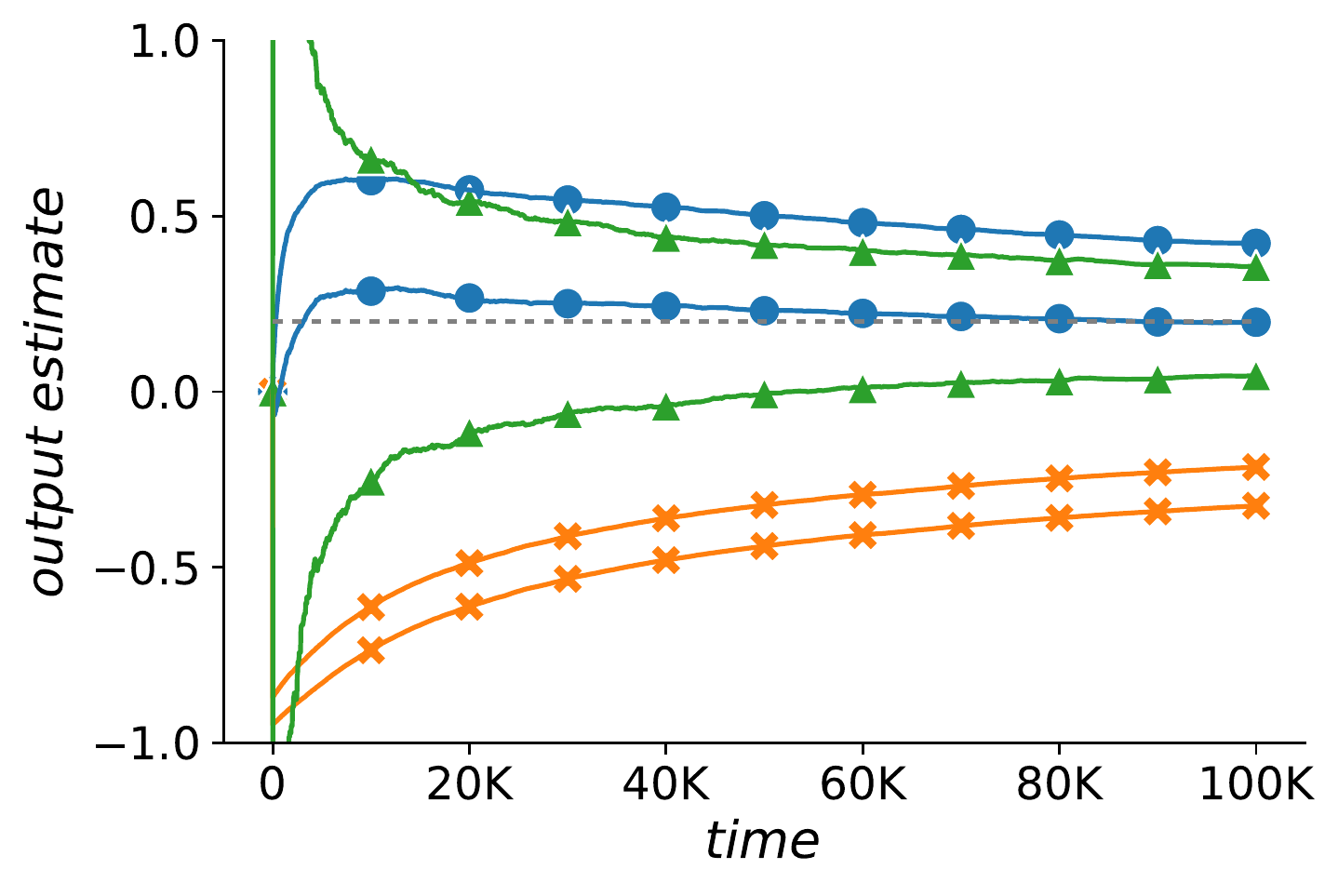}
		\includegraphics[scale=0.275]{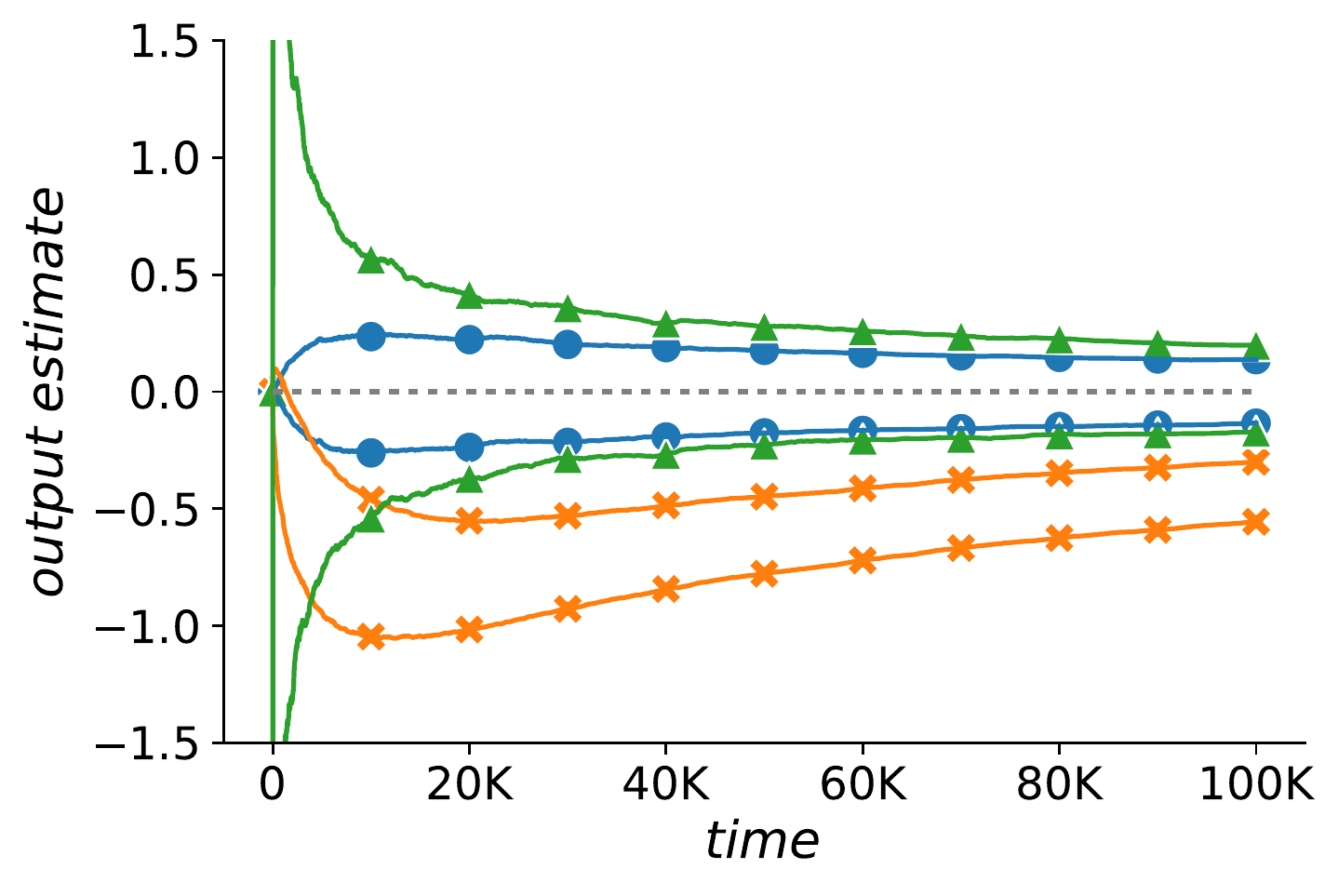}
		\includegraphics[scale=0.275]{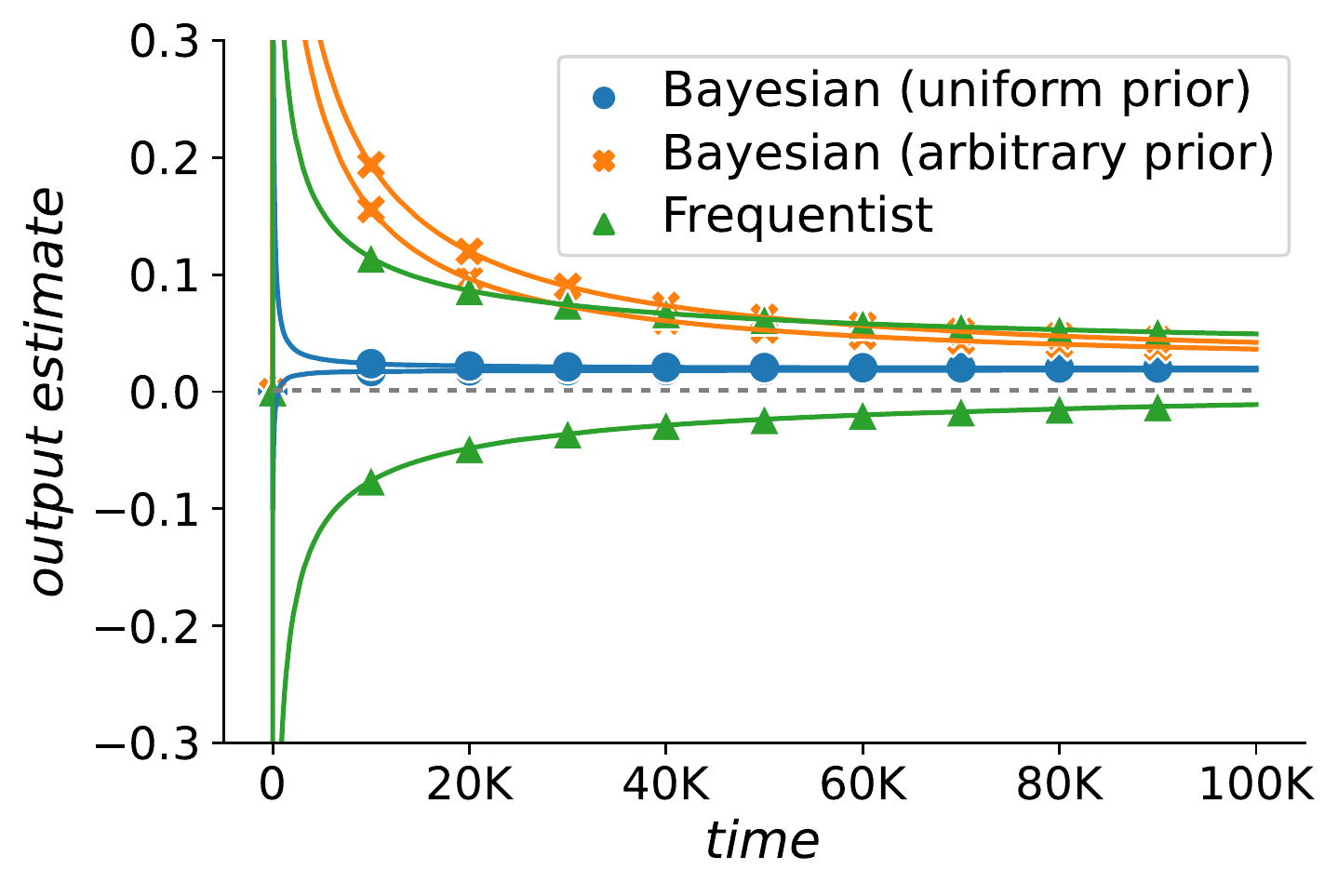}
	\end{minipage}\\
	\begin{minipage}[t]{\textwidth}
		\vspace{0pt}
		\begin{tabular}{|m{4cm}|
					>{\centering\arraybackslash}m{1.5cm}|
					>{\centering\arraybackslash}m{1.5cm}|
					>{\centering\arraybackslash}m{1.5cm}|
					>{\centering\arraybackslash}m{1.5cm}|
					>{\centering\arraybackslash}m{1.5cm}|}
		\hline
		\multirow{2}{*}{Scenario} & \multirow{2}{*}{\shortstack{Size of\\ expression}} & \multicolumn{2}{c|}{Av.\ comp.\ time/step} & \multicolumn{2}{c|}{$\#$ registers}\\
		\cline{3-6}
		& & Freq. & Bayes. & Freq. & Bayes.\\
		\hline
		Lending (bias) + dem.\ par. & $1$ & \num{13.0}\si{\us}  & \num{29.3}\si{\us} & $15$ & $17$\\
		\hline
		Lending (fair) + eq.\ opp. & $5$	&	\num{21.6}\si{\us} & \num{31.0}\si{\us} & $29$ & $27$ \\
		\hline
		Admission + soc.\ burden & $19$ & \num{53.8}\si{\us} & \num{184.6}\si{\us} & $46$ & $102$ \\
		\hline
	\end{tabular}
	\end{minipage}
\end{figure}

We summarize the experimental results in Fig~\ref{fig:experiments}, and, from the table, observe that both monitors are extremely lightweight: they take less than a millisecond per update and small numbers of registers to operate.
From the plots, we observe that the frequentist monitors' outputs are always centered around the ground truth values of the properties, empirically showing that they are always objectively correct.
On the other hand, the Bayesian monitors' outputs can vary drastically for different choices of the prior, empirically showing that the correctness of outputs is subjective.
It may be misleading that the outputs of the Bayesian monitors are wrong as they often do not contain the ground truth values.
We reiterate that from the Bayesian perspective, the ground truth does not exist.
Instead, we only have a probability distribution over the true values that gets updated after observing the generated sequence of events.
The choice of the type of monitor ultimately depends on the application requirements.

\section{Conclusion}\label{sec:conclusion}
We showed how to monitor algorithmic fairness properties on a Markov chain with unknown transition probabilities.
Two separate algorithms are presented, using the frequentist and the Bayesian approaches to statistics.
The performances of both approaches are demonstrated, both theoretically and empirically.

Several future directions exist.
Firstly, more expressive classes of properties need to be investigated to cover a broader range of algorithmic fairness criteria.
We believe that boolean logical connectives, as well as min and max operators can be incorporated straightforwardly using ideas from the related literature \cite{albarghouthi2019fairness}.
This also adds support for absolute values, since $|x| = \max\set{x,-x}$.
On the other hand, properties that require estimating how often a state is visited would require more information about the dynamics of the Markov chain, including its mixing time. 
Monitoring statistical hyperproperties \cite{dimitrova2020probhyper} is another important direction, which will allow us to encode individual fairness properties \cite{dwork2012fairness}.
Secondly, more liberal assumptions on the system model will be crucial for certain practical applications.
In particular, hidden Markov models, time-inhomogeneous Markov models, Markov decision processes, etc., are examples of system models with widespread use in real-world applications.
Finally, better error bounds tailored for specific algorithmic fairness properties can be developed through a deeper mathematical analysis of the underlying statistics, which will sharpen the conservative bounds obtained through off-the-shelf concentration inequalities.


	\bibliographystyle{splncs04}
	\bibliography{references}
	
	\ifarxiv
	\begin{appendix}

\section{Appendices}
\subsection{Probability Spaces and Random Variables}
\label{app:random variables}
A probability space is a triple $\tup{\Omega,\mathcal{F}_\Omega,\pr_\Omega}$, where 
$\Omega$ is a set of outcomes,
$\mathcal{F}_\Omega\subseteq 2^\Omega$ is a $\sigma$-algebra over the set $\Omega$, and
$\pr_\Omega\colon \mathcal{F}\to [0,1]$ is a probability measure over $\mathcal{F}_\Omega$.
A ($(S,\mathcal{F}_S)$-valued) random variable is a measurable function $X\colon (\Omega,\mathcal{F}_\Omega) \to (S,\mathcal{F}_S)$ where $\mathcal{F}_S$ is a $\sigma$-algebra on the co-domain $S$.
An \emph{observation} of the random variable $X$ is an element $x$ in $\mathcal{F}_S$, which will always be written using the lower-case font of the same character used to denote the random variable. 
The random variable $X$ induces a probability measure on the space $(S,\mathcal{F}_S)$ as follows:
For every $A\in \mathcal{F}_S$, $\pr(X\in A) = \pr_\Omega\left(\set{\omega\in\Omega\mid  X(\omega)\in A}\right)$.
Often we discuss probability measure using a random variable $X$ directly over the space $(S,\mathcal{F}_S)$ without mentioning the underlying probability space $\tup{\Omega,\mathcal{F}_\Omega,\pr_\Omega}$.
Moreover, we often simplify notation by writing $\pr(A)$ instead of $\pr(X\in A)$, if the random variable $X$ is clear from the context.
For a given random variable $X$, we often interchangably use the mean $\mu_X$ and the expected value $\expe(X)$ of $X$.


For a given set $S$, define $\distr(S)$ as the set of every random variable---called a \emph{probability distribution}\footnote{In literature, a probability distribution over $S$ is often defined as a probability measure over $S$, instead of the random variable like us that induces the measure.}%
---with domain $(S,2^S)$. 

Let $X,Y\in \distr(\mathbb{R})$ be a pair of real valued random variables defined using the same underlying probability space $\tup{\Omega,\mathcal{F}_\Omega,\pr_\Omega}$.
For $\odot\in\set{+,-,*,\div}$, define $Z = X\odot Y$ as the random variable $Z\colon (\Omega,\mathcal{F}_\Omega)\to (\mathbb{R},2^{\mathbb{R}})$ such that for every $A\in 2^{\mathbb{R}}$, $\pr(Z\in A) \coloneqq \pr_\Omega\left( \set{\omega\in \Omega \mid X(\omega)\odot Y(\omega)\in A} \right)$, provided $X(\omega)\odot Y(\omega)$ is defined for every $\omega$.

Let $X,Y\in \distr(\mathbb{R})$ be a pair of real valued random variables defined using the same underlying probability space $\tup{\Omega,\mathcal{F}_\Omega,\pr_\Omega}$.
Suppose $X\colon (\Omega,\mathcal{F}_\Omega)\to (S_X,\mathcal{F}_X)$ and $Y\colon (\Omega,\mathcal{F}_\Omega)\to (S_Y,\mathcal{F}_Y)$, and denote $\pr_X$ and $\pr_Y$ as the probability measures induced by the random variables $X$ and $Y$ respectively.
The joint probability measure $\pr_{X,Y}(\cdot)$ of $\pr_X$ and $\pr_Y$ is the unique probability measure on the space $(S_X\times S_Y,\mathcal{F}_X\otimes\mathcal{F}_Y)$, defined as:
\begin{align*}
	\pr_{X,Y}(E) = \pr_\Omega((X,Y)\in E).
\end{align*}

\subsection{Frequentist Monitor: Additional Technical Preliminaries and Detailed Proofs}

\subsubsection{Preliminaries: Confidence Intervals}\label{app:confidence intervals}

We summarize three well-known statistical concentration inequalities.

\begin{theorem}[Hoeffding's inequality]\label{thm:hoeffding}
	Suppose $\seq{X}=X_0,\ldots,X_n$ is a sequence of i.i.d.\ random variables such that every $X_i\in [a,b]$ almost surely, and every $X_i$ has mean $\mu_X$.
	Let $\widehat{\mu}(\seq{X})=\frac{\sum_{k=1}^n X_k}{n}$ be the sample mean (which is a random variable), and $\varepsilon\in \mathbb{R}$ be a constant.
	Then the following holds:
	\begin{equation}
		\pr(|\mu_X-\widehat{\mu}(\seq{X})| \leq \varepsilon) \geq 1-2e^{-\frac{2\varepsilon^2n}{(b-a)^2}}.
	\end{equation}
\end{theorem}

Hoeffding's inequality requires a bound $[a,b]$ on the values of the random variable, which is overcome by the Chebyshev's inequality.

\begin{theorem}[Chebyshev's inequality]\label{thm:chebyshev with true variance}
	Suppose $X$ is a random variable with finite mean $\mu_X$ and finite non-zero variance $\sigma^2$.
	For any given $\varepsilon\in \mathbb{R}$, the following holds:
	\begin{equation}
		\pr(|X-\mu_X|\leq \varepsilon) \geq 1 - \frac{\sigma^2}{\varepsilon^2}.
	\end{equation}
\end{theorem}

\subsubsection{Detailed Proof of Technical Claims}
\label{app:freq:proofs}

%

\medskip
\noindent\textbf{Proof of Thm.~\ref{thm:frequentist:soundness}.}
%
	Fix a problem instance $\tup{\Q,\varphi,\delta}$, and assume that $\varphi$ is division-free so that the routine \algfreqdivfree of Alg.~\ref{alg:frequentist monitor division-free} gives us the desired monitor; we will handle the case with division separately.
	For a given concrete finite path $\seq{x}\in\Q^*$ of the Markov chain, Alg.~\ref{alg:frequentist monitor division-free} computes a sequence $\seq{w}$ (of possibly shorter length), so that if $\seq{x}$ is a concrete sample of the evolution $\seq{X}$ of the Markov chain then $\seq{w}$ is a sample of a sequence $\seq{W}=W_1,W_2,\ldots$ of i.i.d.\ random variables such that $\mu_W = \varphi(M)$.
	Then given the sequence $\seq{w}$, we can estimate the mean $\mu_W$ using the Hoeffding's inequality.
	Consequently, for soundness, we need to prove the claim that 
	(A) every member $W_p$ of $\seq{W}$ has mean $\expe(W_p)=\mu_W = \varphi(M)$, and moreover 
	(B) every pair of $W_p,W_k$ for $p\neq k$ are i.i.d.
	The proof is inductive over the structure of the formula $\varphi$.
	
%

%

\noindent
\textbf{Base case:} If $\varphi$ is a variable $v_{ij}\in V$, then the sequence $\seq{W} $ is the same as a \emph{uniformly random reordering} of the sequence of independent Bernoulli random variables $\seq{Y}^{ij}$: 
that $W_i$-s are Bernoulli follows from Line~\ref{alg:eval:var} of Subr.~$\mathit{Eval}(v_{ij})$, and that the uniformly random reordering happens follows from the invocation of the Subr.~$\mathit{ExtractOutcomes}()$ in Line~\ref{alg:eval:extract} of $\mathit{Eval}(v_{ij})$.
Since $\seq{Y}^{ij}$ is i.i.d., hence a uniform random reordering of $\seq{Y}^{ij}$ is also i.i.d.\ with the same distribution.
On the other hand, if $\varphi$ is a constant $c\in \mathbb{R}$, then for every $p$, $\pr(W_p=c)=1$ (Line~\ref{alg:eval:const} of Subr.~$\mathit{Eval}$).
It follows that both (A) and (B) hold in both cases.

\noindent
\textbf{Induction hypothesis:}
If $\varphi$ is neither a variable nor a constant, then it is of the form $\varphi\equiv\varphi_1\odot\varphi_2$ (recall that $\varphi$ is assumed to be division-free), where $\varphi_1$ and $\varphi_2$ are two subformulas over variables $V_1$ and $V_2$ respectively (with $V_1\cup V_2\subseteq V_{\varphi}$) and $\odot\in \set{+,-,\cdot}$.
Suppose $\seq{U} = U_1,U_2,\ldots$ and $\seq{R} = R_1,R_1,\ldots$ are sequences of independent random variables internally generated by Alg.~\ref{alg:frequentist monitor division-free} for the subformulas $\varphi_1$ and $\varphi_2$ respectively.
Let (A) and (B) hold for both $\seq{U}$ and $\seq{R}$.

\noindent
\textbf{Induction step:}
Given $\varphi=\varphi_1\odot\varphi_2$ as defined above, we have the following possibilities:
\begin{description}
	\item[Case $\varphi\equiv \varphi_1 + \varphi_2$:] It follows from Line~\ref{alg:eval:+} of Subr.~$\mathit{Eval}$ that for every $p$, we have $W_p \coloneqq U_p + R_p$.
		Then using linearity of expectation it follows that $\expe(W_p) = \expe(U_p+R_p) = \expe(U_p)+\expe(R_p)=\mu_U + \mu_R = \varphi(M)$, i.e., (A) holds.
	\item[Case $\varphi\equiv \varphi_1-\varphi_2$:] It follows from Line~\ref{alg:eval:-} of Subr.~$\mathit{Eval}$ that for every $p$, we have $W_p \coloneqq U_p - R_p$.
		Then using linearity of expectation it follows that $\expe(W_p) = \expe(U_p-R_p) = \expe(U_p)-\expe(R_p)=\mu_U - \mu_R = \varphi(M)$, i.e., (A) holds.
	\item[Case $\varphi\equiv \varphi_1 \cdot \varphi_2$:] We distinguish between two cases:
		\begin{description}
			\item[Independent multiplication:] It follows from Line~\ref{alg:eval:ind *} of Subr.~$\mathit{Eval}$ that if $\depends(R_1)\cap \depends(R_2) = \emptyset$, then for every $p$, we have $W_p \coloneqq U_{p}\cdot R_{p}$.
			Then $\expe(W_p) = \expe(U_p\cdot R_p) = \expe(U_p)\cdot \expe(R_p)=\mu_U \cdot  \mu_R = \varphi(M)$, since $U_p$ and $R_p$ are independent.
			Hence (A) holds.
			\item[Dependent multiplication:] It follows from Line~\ref{alg:eval:dep *} of Subr.~$\mathit{Eval}$ that if $\depends(R_1)\cap \depends(R_2) \neq \emptyset$, then for every $p$, we have $W_p \coloneqq U_{2p}\cdot R_{2p+1}$.
			Then $\expe(W_p) = \expe(U_{2p}\cdot R_{2p+1}) = \expe(U_{2p})\cdot \expe(R_{2p+1})=\mu_U \cdot  \mu_R = \varphi(M)$, since $U_{2p}$ and $R_{2p+1}$ are independent.
			Hence (A) holds.
		\end{description}
\end{description}
	Claim (B) follows in all the above cases because the elements of $\seq{W}$ are all i.i.d.\ as $\seq{U}$ and $\seq{R}$ are i.i.d\ sequences.
	This completes the proof.
	
	Now suppose $\varphi$ has at least one division operator, in which case we will need to use the routine \algfreq from Alg.~\ref{alg:frequentist monitor}.
	After assigning distinct labels to the repeatedly occurring variables in $\varphi$ to form $\varphi^l$, we convert $\varphi^l$ to the form $\varphi_a+\frac{\varphi_b}{\varphi_c}$, where $\varphi_a$, $\varphi_b$, and $\varphi_c$ are division-free.
	We employ the monitors $\monitor_a$, $\monitor_b$, and $\monitor_c$ to estimate the values $\varphi_a(M)$, $\varphi_b(M)$, and $\varphi_c(M)$, respectively, and the correctness of the outputs of the respective monitors follow from the argument made in the first part of this proof for division-free formulas.
	Note that the interval estimates $\varphi_a(M)$, $\varphi_b(M)$, and $\varphi_c(M)$ are each with confidence $\delta/3$.
	The claim that the output of \algfreq  follows from the interval arithmetic used to estimate the interval $[\mu_\verdict\pm \varepsilon_\verdict]$, and the union bound used to estimate the overall estimate which is the sum of the individual estimates and equals $\delta$; details can be found in the paper by Albarghouthi et al.~\cite{albarghouthi2019fairness}.

\medskip\noindent
\textbf{Proof of Thm.~\ref{thm:frequentist:complexity}.}
First, let us assume that the PSE $\varphi$ is division-free, so that effectively \algfreq reduces to \algfreqdivfree.
In this case, the number of registers for $\set{c_{ij}}$, $\set{c_i}$, $\set{r_\varphi}$, $\set{t_{ij}^l}$, $\set{b_{\varphi'}}$ can be at most $\mathcal{O}(n)$, where $n$ is the number of terms in the formula $\varphi$.
The total number of registers is dominated by the total space occupied by all the $x_i$-s (each location of the array $x_i$ is interpreted as a register).
We first argue that every $x_i$ can grow up to size at most $\mathcal{O}(n)$.
Moreover, the most amount of registers in $x_i$ are required when the operation involved is a dependent multiplication.
Observe that for every dependent multiplication $\varphi=\varphi_1\cdot\varphi_2$ with $i\in \mathit{Dep}(\varphi_1)\cap \mathit{Dep}(\varphi_2)$, if $\varphi_1$ and $\varphi_2$ need $\mathcal{O}(m_1)$ and $\mathcal{O}(m_2)$ samples of $x_i$, then $\varphi$ needs $m_1+m_2$ samples of $x_i$.
As a result, the size of $x_i$ can be at most $\mathcal{O}(n)$, and hence the total space occupied by all the $x_i$ registers will be $\mathcal{O}(n^2)$.

The transition function of the monitor is implemented by the Subr.~$\mathit{Next}$ and the output function is implemented by the Subr.~$\mathit{UpdEst}$.
The computation time of the transition function is dominated by the $\mathit{Eval}(\varphi^l)$ operation in Line~\ref{alg:freq:next:eval}.
Observe that computation time of $\mathit{Eval}(\cdot)$ is dominated by the computation time of dependent multiplications, where every dependent multiplication requires $\mathcal{O}(n)$ operations to shift every $t_{ij}^l$ by one place (there are $\mathcal{O}(n)$-many $t_{ij}^l$-s).
Thus, in the worst case there will be $\mathcal{O}(n)$ dependent multiplications, giving us the $\mathcal{O}(n^2)$ bound on the computation time.
The Subr.~$\mathit{UpdEst}$ requires constant amount of memory and runs in constant time, which can be easily observed from the pseudocode, giving us the overall quadratic bounds on the computation time and memory.
This proves the last part of the theorem.

When $\varphi$ contains division, then first $\varphi$ is converted to the form $\varphi_a+\frac{\varphi_b}{\varphi_c}$, where $\varphi_a$, $\varphi_b$, and $\varphi_c$ are all division-free.
We will argue that if the size of $\varphi$ is $n$, then the sizes of $\varphi_a$ and $\varphi_c$ are each $\mathcal{O}(n2^{\frac{n}{2}})$, and the size of $\varphi_b$ is $\mathcal{O}(n^22^n)$.
Therefore the computation in \algfreq will be dominated by the invocation of \algfreqdivfree on the sub-expression $\varphi_b$ (Line~\ref{line:alg:freq:phi_c} of Alg.~\ref{alg:frequentist monitor}).
First, observe that any arbitrary PSE $\varphi$ can be translated into a semantically equivalent polynomial PSE $\varphi'$ of size $\mathcal{O}(n2^{\frac{n}{2}})$; a formal treatment of this claim can be found in Lem.~\ref{lemma:nf_blowup}.
Given the polynomial PSE $\varphi'$, we can collect all the division-free monomials as a sum of monomials and use it as our $\varphi_a$, whose size will be at most the size of $\varphi'$, which is $\mathcal{O}(n2^\frac{n}{2})$.
The rest of the monomials of $\varphi'$, the ones which contain divisions, have only single variables in the denominator (because of the syntax of PSEs).
Hence, when we combine them in the form of a single ratio $\frac{\varphi_b}{\varphi_c}$, the denominator $\varphi_c$ is a single monomial, whose size can be at most the size of the PSE $\varphi'$, which is $\mathcal{O}(n2^{\frac{n}{2}})$.
The numerator $\varphi_b$, on the other hand, is a sum (or difference) of $\mathcal{O}(n2^{\frac{n}{2}})$-many monomials, and every monomial can be at most $\mathcal{O}(n2^{\frac{n}{2}})$ large (because in the worst case they are of the form $\varphi_d\cdot \varphi_c$, where $\varphi_d$ is some division-free term and the size of the product can be at most the size of the formula $\varphi'$).
Therefore, the size of $\varphi_b$ can be at most $\mathcal{O}(n^22^n)$, and the invocation of \algfreqdivfree dominates the memory and the computation time in \algfreq.
Since \algfreqdivfree takes $\mathcal{O}(n^2)$ time and $\mathcal{O}(n^2)$ registers for its computation for an input PSE of size $n$, hence, for the input PSE $\varphi_b$ of size $\mathcal{O}(n^22^n)$, \algfreqdivfree would take $\mathcal{O}(n^42^{2n})$ time and $\mathcal{O}(n^42^{2n})$ registers.

\medskip\noindent
\textbf{Proof of Thm.~\ref{thm:frequentist:convergence}.}
Observe that every dependent multiplication requires one additional observation sample from the path, and thus in the worst case we will need $n$ observations ($n$ is the size of the expression) for obtaining one observation of $\seq{W}$. 
On the other hand, from the Hoeffding's inequality, it follows that the minimum number of samples of $\seq{W}$ required for an estimate with error at most $\overline{\varepsilon}$ is:
\begin{equation}	
	 -\frac{(u_\varphi-l_\varphi)^2\ln\left(\frac{\delta}{2}\right)}{2\overline{\varepsilon}^2}.
\end{equation}
Thus the bound follows.

\subsection{Bayesian Monitor: Additional Technical Preliminaries and Detailed Proofs}


\subsubsection{Preliminaries on Bayesian Inference and Conjugate Priors}\label{app:bayes:inference}


Let $\mcset$ be the set of all Markov chains with state space $\Q$ (characterized by the set of transition matrices of size $N\times N$) and initial state $1$. Let be $p\colon \mcset\to \mathbb{R}\cup\{\infty\}$ the prior density function. Suppose we are given a sequence of states $\seq{w}\in \Q^*$. We apply Bayes' theorem to compute the \emph{posterior density function} from the given prior density $p(\cdot)$ as:
\begin{equation*}
	p(\mc\mid \seq{w}) = \frac{\pr(\seq{w}\mid \mc)\cdot p(\mc)}{\pr(\seq{w})} = \frac{\pr(\seq{w}\mid \mc)\cdot p(\mc)}{\int_{\mcset} \pr(\seq{w}\mid \mc')\cdotp(\mc')d\mc'},
\end{equation*}
which forms the core equation in Bayesian inference.

In general, computing the posterior density is quite challenging. However, for certain likelihood function there exist priors for which the posterior distribution belongs to the same distribution family as the prior. 
The \emph{likelihood function} for $\trm \in \mathbb{M}$ is 
\begin{align*}
\likely_{\trm}((c{_{ij}})_{i,j\in\Q}) = \prod_{i=1}^N\prod_{j=1}^N \trm_{ij}^{c_{ij}(\seq{w})}.
\end{align*}
The conjugate prior for this likelihood function is the matrix beta distribution.

\subsubsection{Definition of the Matrix Beta Distribution}\label{app:matrix beta}
The matrix beta distribution $\prd\colon \mcset\to \mathbb{R}\cup \{\infty\}$ is a parameterized probability distribution. Its parameter $\theta$ is an $N\times N$ matrix containing only positive entries. The density is given by $\prd(\mc) = \nconst(\theta)\cdot \likely_{\trm}(\theta-\mathbb{1})$, where $\nconst(\cdot)$ is the so-called \emph{normalization constant} defined as:
\begin{align*}
 \nconst(\theta)\coloneqq \prod_{i=1}^N \frac{\Gamma\left(\sum_{j=1}^N \theta_{ij}\right)}{\prod_{j=1}^N \Gamma(\theta_{ij})},
\end{align*}
with $\Gamma(\cdot)$ being the gamma function \cite[p.~50]{gomez2014b,insua2012bayesian}.
For our purposes it suffices to consider $\theta \in \mathbb{N}^{N\times N}$ as a positive integer matrix.

\subsubsection{Derivation of $\mathcal{H}$ and its Update}\label{app:bayes:h & update}
For convenience we use the following more elaborate notation in the proof. Let $n\in\mathbb{N}$ and let $K\in \mathbb{Z}$ s.t. $n-|K|>0$, we define
$\mfact{+}{K}{n}=\prod_{k=0}^{|K|-1} n+k$ and $\mfact{-}{K}{n}=\prod_{k=0}^{|K|-1} n-(k+1)$. 
Notice that 
\begin{align*}
    \mfact{+}{K}{n} = \Myperm[n+|K|-1]{|K|}  \quad\quad \text{and} \quad \quad  \mfact{-}{K}{n} = \Myperm[n-1]{|K|}
\end{align*}
Moreover, observe that
    \begin{align*}
        \mfact{+}{K}{n} =  \mfact{+}{K}{n} \cdot \frac{n+|K|}{n} \quad \text{and} \quad  \mfact{-}{K}{(n+1)} =  \mfact{-}{K}{n} \cdot \frac{n}{n-|K|}
    \end{align*}
Let $\matr{A}\in \mathbb{N}^{N\times N}$, let $\matr{B} \in \mathbb{Z}^{N\times N}$ such that $\matr{A}+\matr{B}>\matr{0}$, let $\star\in\{+,-,0\}$ and $\circ \in \{+,-\}$ we define
\begin{align*}
    \uifact{\star}{\circ}{\matr{B}}{\matr{A}} \coloneqq
    \prod_{i\in \indi_{\matr{B}}^\star}  \prod_{j \in \matr{B}_{i}^{\circ}} \mfact{\circ}{\matr{B}_{ij}}{\matr{A}_{ij}} \quad \text{and} \quad  \uifact{\star}{}{\matr{B}}{\matr{A}} \coloneqq
    \prod_{i\in \indi_{\matr{B}}^\star} \mfact{\star}{\isum{\matr{B}_{i}}}{\isum{\matr{A}_{i}}}
\end{align*}
Where $\isum{\matr{B}_{i}}$ denotes the sum of the row vector $B_i$. 
To make the derivation cleaner we define $\uf$ slightly different (but equivalent) to Equation \ref{equ:bayes:def H}.
\begin{align*}
    \ufact{\matr{B}}{\matr{A}} \coloneqq \left(\frac{       \prod_{i\in \matr{B}^{-}} \mfact{-}{\isum{\matr{B}_{i}}}{\isum{\matr{A}_{i}}}
    }{
        \prod_{i\in \matr{B}^{+}} \mfact{+}{\isum{\matr{B}_{i}}}{\isum{\matr{A}_{i}}}
    } \right)
    \cdot \left(
        \prod_{i=1}^{N} 
        \frac{
             \prod_{j \in \indj_{\matr{B}}^{+}(i)} \mfact{+}{\matr{B}_{ij}}{\matr{A}_{ij}}
        }{
            \prod_{j \in \indj_{\matr{B}}^{-}(i)} \mfact{-}{\matr{B}_{ij}}{\matr{A}_{ij}}
        }
        \right)
\end{align*}
For any matrix $D \in \mathbb{R}^{N \times N} $ we define $D_i^+\coloneqq \set{j\mid D_{ij}>0}$, $D_i^-\coloneqq \set{j\mid D_{ij}<0}$, $D^+\coloneqq \set{i\mid \isum{D_i}>0}$, and $D^-\coloneqq \set{i\mid \isum{D_i}<0}$.
 Additionally, let $D_i^-\coloneqq \set{j\mid D_{ij}=0}$ and 
 $D^-\coloneqq \set{j\mid \isum{D_i}=0}$

\begin{lemma}
    \label{lemma:reduce}
    Let $\matr{A} \in \mathbb{N}^{N\times N}$, let $\matr{B} \in \mathbb{Z}^{N\times N}$ such that $\matr{A}+\matr{B}>\matr{0}$. 
    \begin{align*}
        \nconst(\matr{A}+\matr{B})=\nconst(\matr{A})\cdot \ufact{\matr{B}}{\matr{A}}^{-1}
    \end{align*}
\end{lemma}
\begin{proof}
    From the definition we know that 
    \begin{align*}
        \nconst(\matr{A}+\matr{B})= \prod_{i=1}^{N}\frac{\Gamma\left(\sum_{j=1}^{N} \matr{A}_{ij} + \matr{B}_{ij}\right)}{\prod_{j=1}^{N}\Gamma(\matr{A}_{ij} + \matr{B}_{ij})}
    \end{align*}
   We can split $\nconst(\matr{A}+\matr{B})$ into
    \begin{align*}
        &\left(\prod_{i\in \matr{B}^{+}} \Gamma\left(\sum_{j=1}^{N} \matr{A}_{ij} + \matr{B}_{ij}\right) \right)  \cdot \left(\prod_{i\in \matr{B}^{+}} \prod_{i=1}^{N}\Gamma(\matr{A}_{ij} + \matr{B}_{ij}) \right)^{-1} \\
        &  \cdot       \left(\prod_{i\in \matr{B}^{-}} \Gamma\left(\sum_{j=1}^{N} \matr{A}_{ij} + \matr{B}_{ij}\right) \right)  \cdot \left(\prod_{i\in \matr{B}^{-}} \prod_{i=1}^{N}\Gamma(\matr{A}_{ij} + \matr{B}_{ij}) \right)^{-1} \\ \\
        &    \cdot      \left(\prod_{i\in \matr{B}^{0}} \Gamma\left(\sum_{j=1}^{N} \matr{A}_{ij} + \matr{B}_{ij}\right) \right)  \cdot \left(\prod_{i\in \matr{B}^{0}} \prod_{i=1}^{N}\Gamma(\matr{A}_{ij} + \matr{B}_{ij}) \right)^{-1} \\
    \end{align*}
    We know that for all $n\in \mathbb{N}$ the recursion $\Gamma(n+1) = n \Gamma(n)$ holds. We observe that 
    \begin{align*}
        \prod_{i\in \matr{B}^{+}} \Gamma\left(\sum_{j=1}^{N} \matr{A}_{ij} + \matr{B}_{ij}\right) 
        &=  \left(\prod_{i\in \matr{B}^{+}} \Gamma\left( \isum{\matr{A}_i} \right) \right) \cdot \left(\prod_{i\in \matr{B}^{+}}  \prod_{k=0}^{\isum{\matr{B}_i}-1}  \left(\isum{\matr{A}_i}  + k \right) \right)
    \end{align*}
    and 
    \begin{align*}
        \prod_{i\in \matr{B}^{-}} \Gamma\left(\sum_{j=1}^{N} \matr{A}_{ij} + \matr{B}_{ij}\right) 
        &=  \left(\prod_{i\in \matr{B}^{-}} \Gamma\left( \isum{\matr{A}_i} \right) \right) \cdot \left(\prod_{i\in \matr{B}^{-}}  \prod_{k=0}^{\isum{\matr{B}_i}-1}  \left(\isum{\matr{A}_i}  - k -1 \right) \right)^{-1}
    \end{align*}
    and 
    \begin{align*}
        \prod_{i\in \matr{B}^{0}} \Gamma\left(\sum_{j=1}^{N} \matr{A}_{ij} + \matr{B}_{ij}\right) 
        &=  \prod_{i\in \matr{B}^{0}} \Gamma\left( \isum{\matr{A}_i}\right)
    \end{align*}
    Hence, we obtain 
    \begin{align*}
        &\left(\prod_{i\in \matr{B}^{+}} \Gamma\left( \isum{\matr{A}_i}\right) \right) 
        \cdot
         \uifact{+}{}{\matr{B}}{\matr{A}}
         \cdot 
         \left(\prod_{i\in \matr{B}^{+}} \prod_{i=1}^{N}\Gamma(\matr{A}_{ij} + \matr{B}_{ij}) \right)^{-1} \\
         &\left(\prod_{i\in \matr{B}^{-}} \Gamma\left( \isum{\matr{A}_i}\right) \right) 
         \cdot
         \frac{1}{\uifact{-}{}{\matr{B}}{\matr{A}}}
          \cdot 
          \left(\prod_{i\in \matr{B}^{-}} \prod_{i=1}^{N}\Gamma(\matr{A}_{ij} + \matr{B}_{ij}) \right)^{-1} \\
          &\left(\prod_{i\in \indi_{\matr{Bx}}^0} \Gamma\left( \isum{\matr{A}_i}\right) \right)  \cdot
           \left(\prod_{i\in \matr{B}^{0}} \prod_{i=1}^{N}\Gamma(\matr{A}_{ij} + \matr{B}_{ij}) \right)^{-1} \\
    \end{align*}
    For any $i \in \iinter{0}{N}$ we can split $\prod_{i=1}^{N}\Gamma(\matr{A}_{ij} + \matr{B}_{ij})$ into 
    \begin{align*}
       \left( \prod_{j\in \indj_{\matr{B}}^{+}(i)} \Gamma(\matr{A}_{ij} + \matr{B}_{ij}) \right) \cdot  \left( \prod_{j\in \matr{B}_{i}^{-}}  \Gamma(\matr{A}_{ij} + \matr{B}_{ij}) \right) \cdot  \left( \prod_{j\in \matr{B}_{i}^{0}} \Gamma(\matr{A}_{ij} + \matr{B}_{ij}) \right)
    \end{align*}
    Now notice,
    \begin{align*}
        \prod_{j\in \matr{B}_{i}^{+}}  \Gamma(\matr{A}_{ij} + \matr{B}_{ij}) &= 
       \prod_{j\in \matr{B}_{i}^{+}}  \Gamma(\matr{A}_{ij}) \prod_{k=0}^{\matr{B}_{ij}-1} \left(\matr{A}_{ij} + k\right) \\
       &=\left(\prod_{j\in \matr{B}_{i}^{+}}  \Gamma(\matr{A}_{ij}) \right)\cdot  \left( \prod_{j\in \matr{B}_{i}^{+}} \prod_{k=0}^{\matr{B}_{ij}-1} \left(\matr{A}_{ij} + k\right)\right)
    \end{align*}
    and 
    \begin{align*}
     \prod_{j\in \matr{B}_{i}^{-}}  \Gamma(\matr{A}_{ij} + \matr{B}_{ij})  &= 
        \prod_{j\in \matr{B}_{i}^{-}}  \Gamma(\matr{A}_{ij}) \left(\prod_{k=0}^{|\matr{B}_{ij}|-1} \left(\matr{A}_{ij} - k-1\right)\right)^{-1} \\
        &= 
       \left( \prod_{j\in \matr{B}_{i}^{-}}  \Gamma(\matr{A}_{ij})\right)\cdot  \left(\prod_{j\in \matr{B}_{i}^{-}}  \prod_{k=0}^{|\matr{B}_{ij}|-1} \left(\matr{A}_{ij} - k-1\right)\right)^{-1}
    \end{align*}
    Hence, $\prod_{i=1}^{N}\Gamma(\matr{A}_{ij} + \matr{B}_{ij})$ is equivalent to 
    \begin{align*}
        \left(\prod_{i=1}^{N} \Gamma(\matr{A}_{ij}) \right) \cdot \left( \prod_{j\in \matr{B}_{i}^{+}} \prod_{k=0}^{\matr{B}_{ij}-1} \left(\matr{A}_{ij} + k\right)\right) \cdot  \left(\prod_{j\in \matr{B}_{i}^{-}}  \prod_{k=0}^{|\matr{B}_{ij}|-1} \left(\matr{A}_{ij} - k-1\right)\right)^{-1}
    \end{align*}
    Combining everything we obtain 
    \begin{align*}
        &\left(\prod_{i\in \matr{B}^{+}} \Gamma\left( \isum{\matr{A}_i}\right) \right) 
        \cdot
         \uifact{+}{}{\matr{B}}{\matr{A}}
         \cdot 
         \left(\prod_{i\in \matr{B}^{+}} \prod_{i=1}^{N} \Gamma(\matr{A}_{ij}) \right)^{-1} \cdot \frac{\uifact{+}{-}{\matr{B}}{\matr{A}}}{\uifact{+}{+}{\matr{B}}{\matr{A}}}\\
         &\left(\prod_{i\in \matr{B}^{-}} \Gamma\left( \isum{\matr{A}_i}\right) \right) 
         \cdot
         \frac{1}{\uifact{-}{}{\matr{B}}{\matr{A}}}
          \cdot 
          \left(\prod_{i\in \matr{B}^{-}} \prod_{i=1}^{N} \Gamma(\matr{A}_{ij}) \right)^{-1} \cdot \frac{\uifact{-}{-}{\matr{B}}{\matr{A}}}{\uifact{-}{+}{\matr{B}}{\matr{A}}} \\
          &\left(\prod_{i\in \matr{B}^{0}} \Gamma\left( \isum{\matr{A}_i}\right) \right) 
          \left(\prod_{i\in \matr{B}^{0}} \prod_{i=1}^{N} \Gamma(\matr{A}_{ij}) \right)^{-1} \cdot \frac{\uifact{0}{-}{\matr{B}}{\matr{A}}}{\uifact{0}{+}{\matr{B}}{\matr{A}}}
    \end{align*}
    Now by rearranging the products we obtain 
    \begin{align*}
        c(\matr{A}) \cdot 
         \frac{\uifact{+}{}{\matr{B}}{\matr{A}} \cdot \uifact{+}{-}{\matr{B}}{\matr{A}}}{\uifact{+}{+}{\matr{B}}{\matr{A}}}
         \cdot
         \frac{\uifact{-}{-}{\matr{B}}{\matr{A}}}{\uifact{-}{}{\matr{B}}{\matr{A}}\cdot \uifact{-}{+}{\matr{B}}{\matr{A}}} \cdot
        \frac{\uifact{0}{-}{\matr{B}}{\matr{A}}}{\uifact{0}{+}{\matr{B}}{\matr{A}}}
    \end{align*}
    The rest follows from the definitions of $\uifact{\star}{\circ}{\matr{B}}{\matr{A}}$ and $\uifact{\star}{}{\matr{B}}{\matr{A}}$
\end{proof}

For $i,j \in \Q$, we define $ \matru{i}{j}$ be a $N\times N$ matrix which is $1$ at entry $(i,j)$ and $0$ otherwise.

\begin{lemma}
    \label{lemma:iterate}
    Let $\matr{A} \in \mathbb{N}^{N\times N}$, let $\matr{B}\in \mathbb{Z}^{N\times N}$, for any $a,b\in\iinter{1}{N}$
    \begin{align*}
        \ufact{\matr{B}}{\matr{A}+ \matru{a}{b}} =\ufact{\matr{B}}{\matr{A}} \cdot  \left(\frac{\isum{\matr{A}_a} + \isum{\matr{B}_a}}{\isum{\matr{A}_a}}\right)\cdot\left(\frac{\matr{A}_{ab}}{\matr{A}_{ab} + \matr{B}_{ab}} \right)    
    \end{align*}
\end{lemma}
\begin{proof}
    For $\star\in\{+,-\}$ we observe
    \begin{align*}
        \prod_{i\in \matr{B}^{\star}}\mfact{\star}{\isum{\matr{B}_i}}{\isum{\matr{A}_i+\matru{a}{b}_i}} &= \left(\prod_{i\in \matr{B}^{\star}\setminus \{a\}} \mfact{\star}{\isum{\matr{B}_i}}{\isum{\matr{A}_i}} \right) \cdot \left(\mfact{\star}{\isum{\matr{B}_a}}{(\isum{\matr{A}_a}+1)} \right) \\
        &=
        \left(\prod_{i\in \matr{B}^{\star}} \mfact{\star}{\isum{\matr{B}_i}}{\isum{\matr{A}_i}} \right) \cdot \left(\frac{\isum{\matr{A}_a} +\isum{\matr{B}_a}}{\isum{\matr{A}_a}}\right)^{\star 1} \\
    \end{align*}
    For $\circ\in\{+,-\}$ we observe
\begin{align*}
    \prod_{j\in \indj_{\matr{B}}^{\circ}(a)}\mfact{\circ}{B_{aj}}{(A_{aj}+\matru{a}{b}_{aj})} &= \left(\prod_{j\in \indj_{\matr{B}}^{\circ}(a)\setminus \{b\}} \mfact{\circ}{B_{aj}}{A_{aj}} \right) \cdot \left(\mfact{\circ}{\matr{B}_{ab}}{(\matr{A}_{ab}+1)} \right) \\
    &=
    \left(\prod_{j\in \indj_{\matr{B}}^{\circ}(a)} \mfact{\circ}{B_{aj}}{A_{aj}} \right) \cdot \left(\frac{\matr{A}_{ab} +\matr{B}_{ab}}{\matr{A}_{ab}}\right)^{\circ 1} \\
\end{align*}

Consider the values of $\ufact{\matr{B}}{\matr{A}+\matru{a}{b}}$.
If $\star=+$. If $\circ=+$ then
\begin{align*}
     &\frac{
        \uifact{+}{+}{\matr{B}}{\matr{A}}\cdot \left(\frac{\matr{A}_{ab} +\matr{B}_{ab}}{\matr{A}_{ab}}\right)
        }{
            \uifact{+}{}{\matr{B}}{\matr{A}}\cdot \uifact{+}{-}{\matr{B}}{\matr{A}} \cdot \left(\frac{\isum{\matr{A}_a} +\isum{\matr{B}_a}}{\isum{\matr{A}_a}}\right)}
         \cdot 
         \frac{
        \uifact{-}{}{\matr{B}}{\matr{A}} \cdot \uifact{-}{+}{\matr{B}}{\matr{A}}
        }{
            \uifact{-}{-}{\matr{B}}{\matr{A}}} 
        \cdot 
        \frac{
        \uifact{0}{+}{\matr{B}}{\matr{A}}
        }{
            \uifact{0}{-}{\matr{B}}{\matr{A}}} \\
            &=\ufact{\matr{B}}{\matr{A}+\matru{a}{b}} \cdot \frac{\matr{A}_{ab} +\matr{B}_{ab}}{\matr{A}_{ab}} \cdot \frac{\isum{\matr{A}_a}}{\isum{\matr{A}_a} +\isum{\matr{B}_a}}
\end{align*}
If $\circ=-$ then
\begin{align*}
    &\frac{
       \uifact{+}{+}{\matr{B}}{\matr{A}}
       }{
           \uifact{+}{}{\matr{B}}{\matr{A}}\cdot \uifact{+}{-}{\matr{B}}{\matr{A}} \cdot \left(\frac{\isum{\matr{A}_a} +\isum{\matr{B}_a}}{\isum{\matr{A}_a}}\right) \cdot \left(\frac{\matr{A}_{ab}}{\matr{A}_{ab} +\matr{B}_{ab}}\right)}
        \cdot 
        \frac{
       \uifact{-}{}{\matr{B}}{\matr{A}} \cdot \uifact{-}{+}{\matr{B}}{\matr{A}}
       }{
           \uifact{-}{-}{\matr{B}}{\matr{A}}} 
       \cdot 
       \frac{
       \uifact{0}{+}{\matr{B}}{\matr{A}}
       }{
           \uifact{0}{-}{\matr{B}}{\matr{A}}} \\
           &=\ufact{\matr{B}}{\matr{A}+\matru{a}{b}} \cdot \frac{\matr{A}_{ab} +\matr{B}_{ab}}{\matr{A}_{ab}} \cdot \frac{\isum{\matr{A}_a}}{\isum{\matr{A}_a} +\isum{\matr{B}_a}}
\end{align*}
If $\star=-$. If $\circ=+$ then
\begin{align*}
     &\frac{
        \uifact{+}{+}{\matr{B}}{\matr{A}}
        }{
            \uifact{+}{}{\matr{B}}{\matr{A}}\cdot \uifact{+}{-}{\matr{B}}{\matr{A}}}
         \cdot 
         \frac{
        \uifact{-}{}{\matr{B}}{\matr{A}} \cdot \uifact{-}{+}{\matr{B}}{\matr{A}} \cdot \left(\frac{\isum{\matr{A}_a}}{\isum{\matr{A}_a} +\isum{\matr{B}_a}}\right)\cdot \left(\frac{\matr{A}_{ab} +\matr{B}_{ab}}{\matr{A}_{ab}}\right)
        }{
            \uifact{-}{-}{\matr{B}}{\matr{A}}} 
        \cdot 
        \frac{
        \uifact{0}{+}{\matr{B}}{\matr{A}}
        }{
            \uifact{0}{-}{\matr{B}}{\matr{A}}} \\
            &=\ufact{\matr{B}}{\matr{A}+\matru{a}{b}} \cdot \frac{\matr{A}_{ab} +\matr{B}_{ab}}{\matr{A}_{ab}} \cdot \frac{\isum{\matr{A}_a}}{\isum{\matr{A}_a} +\isum{\matr{B}_a}}
\end{align*}
If $\circ=-$ then
\begin{align*}
    &\frac{
       \uifact{+}{+}{\matr{B}}{\matr{A}}
       }{
           \uifact{+}{}{\matr{B}}{\matr{A}}\cdot \uifact{+}{-}{\matr{B}}{\matr{A}} }
        \cdot 
        \frac{
       \uifact{-}{}{\matr{B}}{\matr{A}} \cdot \uifact{-}{+}{\matr{B}}{\matr{A}}\cdot \left(\frac{\isum{\matr{A}_a}}{\isum{\matr{A}_a} +\isum{\matr{B}_a}}\right)
       }{
           \uifact{-}{-}{\matr{B}}{\matr{A}}\cdot \left(\frac{\matr{A}_{ab}}{\matr{A}_{ab} +\matr{B}_{ab}}\right)} 
       \cdot 
       \frac{
       \uifact{0}{+}{\matr{B}}{\matr{A}}
       }{
           \uifact{0}{-}{\matr{B}}{\matr{A}}} \\
           &=\ufact{\matr{B}}{\matr{A}+\matru{a}{b}} \cdot \frac{\matr{A}_{ab} +\matr{B}_{ab}}{\matr{A}_{ab}} \cdot \frac{\isum{\matr{A}_a}}{\isum{\matr{A}_a} +\isum{\matr{B}_a}}
\end{align*}
If $\star=0$. If $\circ=+$ then
\begin{align*}
     &\frac{
        \uifact{+}{+}{\matr{B}}{\matr{A}}
        }{
            \uifact{+}{}{\matr{B}}{\matr{A}}\cdot \uifact{+}{-}{\matr{B}}{\matr{A}}}
         \cdot 
         \frac{
        \uifact{-}{}{\matr{B}}{\matr{A}} \cdot \uifact{-}{+}{\matr{B}}{\matr{A}} 
        }{
            \uifact{-}{-}{\matr{B}}{\matr{A}}} 
        \cdot 
        \frac{
        \uifact{0}{+}{\matr{B}}{\matr{A}} \cdot \left(\frac{\matr{A}_{ab} +\matr{B}_{ab}}{\matr{A}_{ab}}\right)
        }{
            \uifact{0}{-}{\matr{B}}{\matr{A}}} \\
            &=\ufact{\matr{B}}{\matr{A}+\matru{a}{b}} \cdot \frac{\matr{A}_{ab} +\matr{B}_{ab}}{\matr{A}_{ab}} 
\end{align*}
If $\circ=-$ then
\begin{align*}
    &\frac{
       \uifact{+}{+}{\matr{B}}{\matr{A}}
       }{
           \uifact{+}{}{\matr{B}}{\matr{A}}\cdot \uifact{+}{-}{\matr{B}}{\matr{A}} }
        \cdot 
        \frac{
       \uifact{-}{}{\matr{B}}{\matr{A}} \cdot \uifact{-}{+}{\matr{B}}{\matr{A}}
       }{
           \uifact{-}{-}{\matr{B}}{\matr{A}}} 
       \cdot 
       \frac{
       \uifact{0}{+}{\matr{B}}{\matr{A}}
       }{
           \uifact{0}{-}{\matr{B}}{\matr{A}}}\cdot \left(\frac{\matr{A}_{ab}}{\matr{A}_{ab} +\matr{B}_{ab}}\right) \\
           &=\ufact{\matr{B}}{\matr{A}+\matru{a}{b}} \cdot \frac{\matr{A}_{ab} +\matr{B}_{ab}}{\matr{A}_{ab}} 
\end{align*}

\end{proof}

\subsubsection{Polynomial Form}

\begin{lemma}
    Any PSE containing only divisions of the form $\frac{1}{\lp{i}{j}}$ can be transformed into a polynomial.
\end{lemma}
\begin{proof}
    Let $\varphi$ and $\varphi'$ be two polynomials.
    Then $\varphi + \psi $ is a polynomial, i.e.\
    \begin{align*}
        \varphi + \varphi' &= \sum_{k=1}^p \kappa_k\prod_{i=1,j=1}^{N} \lp{i}{j}^{d_{ij}^k} + \sum_{l=1}^q \kappa_l'\prod_{i=1,j=1}^{N} \lp{i}{j}^{{d'}_{ij}^l}
    \end{align*}
    Then $\varphi \cdot \psi $ is a polynomial, i.e.\
    \begin{align*}
        \varphi \cdot \varphi' &= \sum_{k=1}^p \kappa_k \prod_{i=1,j=1}^{N} \lp{i}{j}^{d_{ij}^k} \cdot \sum_{l=0}^q \kappa_l'\prod_{i=1,j=1}^{N}  \lp{i}{j}^{{d'}_{ij}^l}\\
        &=\sum_{k=1}^p \sum_{l=1}^q \kappa_k \kappa_l'\prod_{i=1,j=1}^{N} \lp{i}{j}^{d_{ij}^k} \cdot \prod_{i=1,j=1}^{N} \lp{i}{j}^{{d'}_{ij}^l}
    \end{align*}
    Trivially the leafs, i.e. $\lp{i}{j}$ or $1 \div \lp{i}{j}$ of the formula tree are polynomials. Hence, by starting from the leafs and propagating the transformations upwards we obtain a formula in polynomial form.
\end{proof}

\begin{lemma}
    \label{lemma:formula_m}
    Let $m\in\mathbb{N}$ s.t. $m\geq 2$, let 
    \begin{align*}
        \varphi_m\coloneqq \prod_{i=0}^{m-1} (q_{2i} + q_{2i+1})
    \end{align*}
    containing $2m$ unique variables then its polynomial form is of size $2^{2m+1}-1$.
\end{lemma}
\begin{proof}
    For some $m\in \mathbb{N}$, we show by induction that the polynomial form of $\varphi_m$ is 
$\sum_{i=0}^{2^{m}-1} \prod_{j=0}^{m-1} q_{x_{ij}}$
    where $x_{ij} \in [0..m]$. First,
    \begin{align*}
        \varphi_2= (q_0+q_1)\cdot (q_2+q_3) = q_0q_2+q_0q_3+q_1q_2+q_1q_3 =  \sum_{i=0}^{2^2-1} \prod_{j=0}^{2-1} q_{x_{ij}}
    \end{align*}
    Second, by IH
    \begin{align*}
        \varphi_{k+1}&= \varphi_{k}\cdot (q_{2k}+q_{2k+1}) =  \left(\sum_{i=0}^{2^{k}} \prod_{j=0}^{k-1} q_{x_{ij}} \right) \cdot (q_{2k}+q_{2k+1}) =  \\
        &=  \sum_{i=0}^{2^{k}-1} q_{2k} \prod_{j=0}^{k-1} q_{x_{ij}}  + \sum_{i=0}^{2^{k}-1} q_{2k+1} \prod_{j=0}^{k-1} q_{x_{ij}} = \sum_{i=0}^{2^k} \prod_{j=0}^k q_{x_{ij}}
    \end{align*}
    Therefore, the sum consists of $2^m-1$ additions symbols and $2^m$ products, with each product containing $m$ variable symbols and $m-1$ product symbols.
    Thus we obtain $2^m(m+m-1)+2^{m}-1 = 2^{m+1}m -1$.
\end{proof}
\begin{lemma}
    \label{lemma:nf_blowup}
    Every PSE $\varphi$ can be reduced to a PSE $\psi$ that is in polynomial form, such that $\varphi$ and $\psi$ are semantically equivalent.
	If the size of $\varphi$ is $n$, then the size of $\psi$ is bounded by $\mathcal{O}\left(n2^{\frac{n}{2}}\right)$.
\end{lemma}
\begin{proof}
    Choose $m$ s.t. $4(m-1)-1\leq n\leq 4m-1$. Then $|\varphi_{m-1}|<\varphi\leq |\varphi_{m}|$ and from Lemma \ref{lemma:formula_m} that $\varphi_{m}$'s polynomial form is smaller than $2^{m+1}m -1$.
\end{proof}

\subsubsection{Soundness of Algorithm \ref{alg:bayesian monitor}}\label{app:bayes:soundness}

To compute the expected value of a PSE $\varphi$ w.r.t.\ to a matrix beta distribution we utilise results about the Dirichlet distribution. For some $N\in \mathbb{N}^+$, $a \in (\mathbb{R}^+)^N$ the probability density function of the Dirichlet distribution is defined as
\begin{align*}
p_a(v) \coloneqq  \frac{\Gamma\left(\sum_{i=1}^N a_{i}\right)}{\prod_{i=1}^N \Gamma(a_{i})} \cdot \prod_{j=1}^N m_i^{a_{i}-1} 
\end{align*}
for any $m \in \simplex(N-1)$.
Moreover, let $b \in \mathbb{R}^N$ then we can compute the expectation \cite{marchal2017sub}
\begin{align*}
    \expe_a\left(\prod_{i=1}^N m_{i}^{b_{i}}\right) =  \frac{\Gamma\left(\sum_{i=1}^N a_{i}\right)}{\prod_{i=1}^N \Gamma(a_{i})} \cdot   \frac{\prod_{i=1}^N \Gamma(a_{i}+b_i)}{\Gamma\left(\sum_{i=1}^N a_{i} + b_i\right)}
\end{align*}
Notice that the probability density function of the matrix beta distribution is the product of $N$ different Dirichlet probability density functions. Hence, this result can easily be extended to matrix beta distributions.

That is, for some set of parameter $\theta$ and some matrix $A \in \mathbb{N}^{N\times N}$ we obtain
\begin{align*}
 \expe_{\theta}\left(\prod_{i=1}^N\prod_{j=1}^N M_{ij}^{A_{ij}}\right) = \frac{\nconst(\theta)}{\nconst(\theta +A)} 
\end{align*}
by independence.
Moreover, notice that if $\theta=\mathbb{1}$, i.e.\ the uniform prior, we obtain 
\begin{align*}
    \int_{\simplex(N,N)} \likely_{\mcm}(A)\;d\mcm = \expe_{1}\left(\prod_{i=1}^N\prod_{j=1}^N M_{ij}^{A_{ij}}\right) = \frac{1}{\nconst(A)+\mathbb{1}} 
   \end{align*}
Moreover, we extend this result further to allow for limited negative powers. 

From now on we use $C=(c_{ij})_{i,j\in\Q}$ for the count matrix and $D=(d_{ij})_{i,j\in\Q}$ for the exponent matrix
of a string $\seq{w}\in\Q$.

\begin{lemma}
    \label{lemma:exp_product}
    Let $\theta$ be matrix beta parameter and let D$\varphi=\kappa_1 \cdot \eta_1$ be a monomial.
    \begin{align*}
        \expe_{\theta}\left(\beval{\varphi}{\mcmr} \mid \seq{w} \right) = \kappa_1 \cdot \ufact{\matr{\mlp}}{\matr{\para} + C}
    \end{align*} 
    if $\matr{\para} + C +D\geq 0$, 
\end{lemma}
\begin{proof}
    By definition and Bayes' Theorem and linearity we obtain 
    \begin{align*}
        \expe_{\theta}\left( \beval{\varphi}{\mcmr}\mid \seq{w} \right) &= \int_{\simplex(N,N)} \beval{\varphi}{\mcm}  \cdot \bpr{\theta}(\mcm \mid \seq{w}) \; d\mcm \\
        &= \int_{\simplex(N,N)} \beval{\varphi}{\mcm}  \cdot \bpr{\theta}(\mcm \mid C) \; d\mcm \\
        &= \int_{\simplex(N,N)} \kappa_1 \cdot\prod_{i,j\in \Q} \mcm_{ij}^{D_{ij}}  \cdot \nconst(\matr{\para}+ C) \cdot \likely_{\mcm}(\matr{\para}+C - \mathbb{1})  \; d\mcm \\
        &=  \kappa_1\cdot \nconst(\matr{\para}+ C) \cdot \int_{\simplex(N,N)} \likely_{\mcm}(\matr{\para}+  D + C- \mathbb{1})  \; d\mcm
    \end{align*}
    Hence we obtain
    \begin{align*}
        \expe_{\theta}\left(\beval{\varphi}{\mcmr} \mid \seq{w} \right)  = \frac{\nconst(\matr{\para}+ C) }{\nconst(\matr{\para} + \matr{\mlp} + C) }
    \end{align*}
From Lemma \ref{lemma:reduce} we obtain 
    \begin{align*}
        \frac{\nconst(\matr{\para}+ C) }{\nconst(\matr{\para} + \matr{\mlp} + C) } = \frac{\nconst(\matr{\para}+ C) }{\nconst(\matr{\para} + C) \cdot \ufact{\matr{\mlp}}{\matr{\para} + C}^{-1}}  = \ufact{\matr{\mlp}}{\matr{\para} + C}
    \end{align*}
    
\end{proof}


The following Theorem demonstrates that we are able to compute the expected value of a probability property. 
\begin{theorem}
    \label{thm:compute}
    Let $\seq{w}\in \Q^* $, let $\matr{\para} \in \mathbb{N}^{N\times N}$ be a matrix beta parameter, and let $\varphi=\sum_{k=1}^{p} \kappa_k  \eta_k$ be a probability expression in polynomial form.
    Then
    \begin{align*}
     \expe_{\theta}\left( \beval{\varphi}{\mcmr}\mid \seq{w} \right) = \sum_{k=1}^{p} \kappa_k \ufact{D^k}{\matr{\para} +C}
    \end{align*}
    if $\forall k\in \iinter{1}{p}.\matr{\para} + C + D^k \geq 0$.
\end{theorem}
\begin{proof}
This follows directly from Lemma \ref{lemma:exp_product} and the linearity of expectation.
\end{proof}

\begin{proposition}[Confidence interval.]\label{prop:bayes:conf interval using chebyshev}
	Given the problem instance $(\Q,\varphi, \delta)$ and the parameter matrix $\theta$ for the prior distribution, and given a sequence of states $\seq{w}\in\Q^*$, the following is a $(1-\delta)\%$ confidence interval of $\varphi(M)$:
	\begin{equation}\label{equ:bayesian:confidence using chebyshev}
		\left[ \expe_\theta(\varphi(\trm)\mid \seq{w}) \pm \sqrt{\frac{\expe_\theta(\varphi^2(\trm)\mid \seq{w})-\expe_\theta(\varphi(\trm)\mid \seq{w})^2}{\delta}} \right].
	\end{equation}
\end{proposition}

\begin{proof}
	A direct consequence of Chebyshev's inequality (Thm.~\ref{thm:chebyshev with true variance}).
\end{proof}

\subsubsection{Resource and Time bounds of Algorithm \ref{alg:bayesian expectation monitor}}

\begin{lemma}
    \label{lemma:b_counter}
    Let $\varphi$ be a PSE in polynomial form containing $p$ monomials. 
    Then Algorithm \ref{alg:bayesian expectation monitor} requires less than  $V_{\varphi} + \dom(V_{\varphi}) + 2p $ counters.
\end{lemma}
\begin{proof}
    We need a counter for each variable that occurs in $\varphi$ they correspond to $c_{ij}$.
    We need a counter for the number of visit to a particular state, this corresponds to $c_i$, the number of which is smaller than $|\dom(V_{\varphi})|$. 
    And we need 2 counter to store each $h^l$ for $l \in \iinter{1}{p}$ resulting in $2p$ counters.
\end{proof}

\begin{lemma}
    \label{lemma:b_speed}
    Let $\varphi$ be a PSE in polynomial form containing $p$ monomials. 
    Then Algorithm \ref{alg:bayesian expectation monitor} requires updates its verdict in $\mathcal{O}(p)$ time. 
\end{lemma}
\begin{proof}
    We need to update $h^l$ for each $l\in\iinter{1}{l}$. To do so at most two additions, two multiplications and two divisions are required. 
    To compute $E$ we need to multiply each $h_l$ with some constant $\kappa_l$ and sum them up. Resulting in $\mathcal{O}(p)$.
\end{proof}

	\end{appendix}
	\else
	\fi
\end{document}